\documentclass[aps,prl,twocolumn,superscriptaddress,floatfix]{revtex4-2}
\usepackage{amsmath,amssymb,graphicx,bm}
\usepackage{slashed}
\usepackage{color}
\usepackage{physics}

\usepackage{hyperref}

\begin{document}

\title{Matrix Density Waves and Fractionally Charged Point Defects in Flavor Weyl Semimetals}

\author{Shantonu Mukherjee}
\email{shantanumukherjeephy@gmail.com}
\affiliation{Department of Physics, Indian Institute of Technology Bombay,
Powai, Mumbai 400076, India}
\author{Hridis K. Pal}
\email{hridis.pal@iitb.ac.in}
\affiliation{Department of Physics, Indian Institute of Technology Bombay,
Powai, Mumbai 400076, India}

\date{\today}

\begin{abstract}
A conventional interaction-driven Weyl density wave is a complex scalar and
supports vortex lines, but no topologically stable point defects. We show that two Weyl flavors
do not merely duplicate this order. Starting from the matrix-valued
internode coherence, a local flavor-symmetric repulsion selects a traceless
adjoint condensate within the density-wave sector, while the fermionic
ground-state energy locks its complex components into the collinear form
$\bm\Delta=\Phi_0\mathbf n e^{i\theta}$. The resulting order-parameter manifold,
$(S^2\times S^1)/\mathbb Z_2$, supports both unit hedgehogs and half-quantum Alice
strings of the same electronic mass that gaps the Weyl fermions. An elementary
hedgehog binds a single normalizable zero mode whose empty and occupied sectors
carry charges $-e/2$ and $+e/2$ at neutrality. An elementary Alice string carries
a single chiral electronic mode despite its $\pi$ phase winding, and transporting
a hedgehog around it reverses the hedgehog winding, $N\rightarrow-N$. Flavor
therefore converts spontaneous translation breaking into a route to
point-defect fractionalization and intertwines the point- and line-defect sectors
of an interaction-generated electronic mass.
\end{abstract}

\maketitle

Interaction-driven symmetry-broken states in Weyl semimetals can have consequences
with no generic counterpart in conventional metals because they develop from chiral
band crossings carrying topological charge. A simple example is the internode
charge-density wave (CDW)~\cite{weiaji,wangzhang,roysau}. The resulting CDW is an
excitonic insulator in which spontaneous chiral-symmetry breaking dynamically
generates a complex fermion mass. The phase of this mass acts as a dynamical axion,
and a vortex of the phase is therefore an axion string carrying a chiral
mode~\cite{wangzhang,youcho,callanharvey}---a property not generic to conventional
CDW defects.

A natural question is how this picture changes when each Weyl node carries an
additional internal flavor. This question is relevant to several classes of systems.
In Weyl systems with weak spin--orbit coupling, physical spin can provide two nearly
identical Weyl flavors with an approximate $\mathrm{SU}(2)$ symmetry.
Spin-group-protected flavor Weyl points proposed in antiferromagnets such as
CoNb$_3$S$_6$ and GdCuSn provide concrete realizations of this
possibility~\cite{flavorweyl}. Higher-charge and multifold nodes, including those
found in CoSi, RhSi, and AlPt, can also be organized in terms of multiflavor Weyl
parent theories in which the flavors are coupled by flavor-mixing, Lorentz-breaking
perturbations~\cite{multifoldtheory,multifoldexp,sanchezrhsi,schroeteralpt,
multiflavorhigher,mukherjee}.
Complementary realizations may be engineered in synthetic platforms such as
ultracold atomic gases, where different hyperfine states provide controllable
flavors~\cite{coldatomweyl}. What qualitatively new interaction-driven physics can
arise from this internal structure?

We address this question using a minimal model consisting of a pair of
opposite-chirality Weyl nodes, each carrying a degenerate $\mathrm{SU}(2)$ flavor
doublet, and focus on the internode CDW phase. To make direct contact with earlier studies and isolate the
effect of the additional flavor degree of freedom, we retain the
Weyl-pseudospin-scalar internode channel that produces the conventional fully
gapped Weyl density wave, while allowing the order parameter to have a general
structure in flavor space. It is therefore a complex $2\times2$ matrix in the flavor space. We show that a local flavor-symmetric repulsion selects the
traceless adjoint channel over the scalar singlet, while the fermionic ground-state
energy locks the general complex adjoint field into the collinear form
$\Delta^a=\Phi_0 n^a e^{i\theta}$. The resulting order-parameter manifold is
$\mathcal M=(S^2\times S^1)/\mathbb Z_2$; it modifies the elementary line defect of the scalar Weyl CDW and introduces topologically stable point defects absent in the scalar case. The elementary line defect is
a half-quantum vortex in which the CDW phase winds by $\pi$ while the flavor
orientation reverses. This Alice axion string carries a single chiral electronic
mode. Its elementary point defect is a unit hedgehog, which binds a single
normalizable zero mode whose empty and occupied sectors carry charges $-e/2$ and
$+e/2$, respectively, at neutrality in the spectrally symmetric limit. Moreover,
transporting the hedgehog around the Alice string reverses its winding, converting
it into an antihedgehog. The two-flavor Weyl problem thus establishes a
minimal electronic setting for interaction-driven point-defect fractionalization and
intertwined defect physics in three dimensions.

\begin{figure}
        \centering
        \includegraphics[width=1\columnwidth]{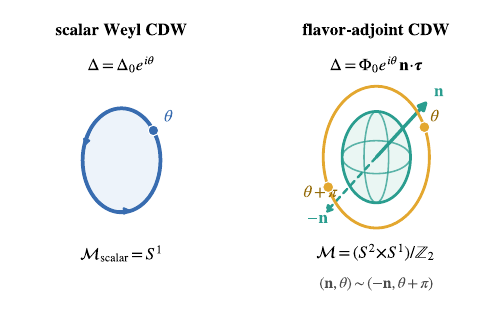}
       \caption{Order-parameter manifolds of the scalar and flavor-adjoint Weyl
CDWs. For a nondegenerate Weyl pair, the internode order is a complex scalar,
$\Delta=\Delta_0e^{i\theta}$, with fixed-amplitude manifold
$\mathcal M_{\rm scalar}=S^1$. For a flavor doublet, the adjoint condensate
$\Delta=\Phi_0e^{i\theta}\mathbf n\cdot\bm\tau$ carries both a CDW phase and an
internal orientation $\mathbf n\in S^2$. The identification
$(\mathbf n,\theta)\sim(-\mathbf n,\theta+\pi)$ gives
$\mathcal M=(S^2\times S^1)/\mathbb Z_2$.}

        \label{fig:1}
    \end{figure}

\textit{Matrix density wave.}---Consider two
opposite-chirality Weyl nodes separated in momentum space by $2\mathbf b$, each carrying two degenerate flavors. The corresponding fields
$\phi_R$ and $\phi_L$ each carry a Weyl-pseudospin and a flavor index. We combine
them into the eight-component spinor $\psi=(\phi_R,\phi_L)^T$, ordered in the
product space
$\mathcal H_\rho\otimes\mathcal H_\sigma\otimes\mathcal H_\tau$.
Here $\bm\rho$, $\bm\sigma$, and $\bm\tau$ denote Pauli matrices acting on the
node, Weyl-pseudospin, and flavor sectors, respectively. Identity matrices and
tensor products between these sectors will henceforth be left implicit. The
low-energy Hamiltonian is
\begin{equation}
H_0=
\int d^3r\,
\psi^\dagger
\left(
-i v_F\rho_z\bm\sigma\cdot\bm\nabla
\right)
\psi.
\label{eq:H0}
\end{equation}
The slowly varying fields enter the microscopic electron operator as
$
c_{\alpha}(\mathbf r)
=
e^{i\mathbf b\cdot\mathbf r}\phi_{R\alpha}(\mathbf r)
+
e^{-i\mathbf b\cdot\mathbf r}\phi_{L\alpha}(\mathbf r),
$
where $\alpha=1,\, 2$ labels flavor. 

We introduce short-range interactions and focus on translation-breaking
internode coherence at wave vector $2\mathbf b$. To isolate the effect of
flavor, we consider the same Weyl-pseudospin-scalar channel that underlies the
conventional fully gapped Weyl density wave~\cite{wangzhang,roysau}, now
allowing a general structure in the additional flavor space. The order
parameter is therefore a general 
$2\times2$ matrix in flavor space:
\begin{equation}
M_{\alpha\beta}
=
\left\langle
\phi_{L\beta}^\dagger\phi_{R\alpha}
\right\rangle
=
\frac{1}{2}
\left(
B_0\tau^0+B_a\tau^a
\right)_{\alpha\beta}.
\label{eq:M}
\end{equation}
The coefficient $B_0$ is the flavor singlet, whereas the three coefficients
$B_a$ form the flavor adjoint.

To determine which channel is favored, consider the local flavor-symmetric
repulsion
$H_U=(U/2)\int d^3r\,{:}n^2{:}$, where $n$ is the total density and $U>0$.
Hartree--Fock decoupling in the full matrix channel gives the quadratic
interaction contribution
\begin{equation}
\delta F_U=
\frac{U}{2}
\left(
|B_0|^2-\sum_a|B_a|^2
\right).
\label{eq:HFselection}
\end{equation}
At the matrix level, the Hartree contribution
$U|\operatorname{Tr}M|^2$ penalizes only the singlet component, whereas the
exchange contribution $-U\operatorname{Tr}(MM^\dagger)$ lowers both channels.
Their combination produces the opposite signs in Eq.~(3). Flavor symmetry makes
the bare fermionic susceptibility identical in the singlet and adjoint
channels. Including the common inverse susceptibility \(r_0\), the singlet and adjoint coefficients become \(r_0+U\) and \(r_0-U\), respectively.
The repulsion therefore hardens the scalar channel and softens the adjoint
channel. As $U$ is increased, the adjoint coefficient reaches zero first, so
the leading instability within this internode density-wave sector is a
flavor-adjoint density wave. The vanishing density of states at a
Weyl node implies a finite critical coupling whose value depends on the
ultraviolet regularization; the relative splitting between the singlet and
adjoint channels follows from flavor symmetry and the local form of the
interaction. Details are given in the Supplemental Material (SM)~\cite{SM}.

The physical character of the adjoint wave is different from the scalar one. The contribution
of the general matrix $M$ to the microscopic total density at wave vector
$2\mathbf b$ is
$\delta n(\mathbf r)
=
2\operatorname{Re}
\left[
e^{2i\mathbf b\cdot\mathbf r}\operatorname{Tr}M
\right].
$
Thus the singlet $B_0$ produces an ordinary scalar CDW. For the pure adjoint
selected above, $\operatorname{Tr}M=0$, and the leading total-density
modulation vanishes. In a flavor basis aligned with the condensate, however,
the two flavor-resolved densities modulate oppositely. The order therefore
still carries momentum $2\mathbf b$ and breaks translation symmetry, but its
modulation resides in flavor space. Moreover, by coherently mixing nodes of
opposite chirality, it generates a Dirac mass and gaps both flavors.

We next project onto the soft adjoint channel. In the representation
$\gamma^0= i\rho_x$, $\gamma^i=i\rho_y\sigma_i$, and
$\gamma^5=\rho_z$, with $\bar\psi=\psi^\dagger\gamma^0$, the effective
adjoint interaction can be written as
\begin{equation}
\mathcal L_{\rm int}
=
G\left[
\left(\bar\psi\tau^a\psi\right)^2
+
\left(i\bar\psi\gamma^5\tau^a\psi\right)^2
\right],
\label{eq:int}
\end{equation}
where $a=1,2,3$ is summed. This is the flavor-adjoint counterpart of the scalar
Nambu--Jona-Lasinio channel used for a conventional Weyl
CDW~\cite{njl,wangzhang}. Hubbard--Stratonovich decoupling introduces the
complex triplet
$
\Delta^a=u^a+i v^a
\propto
\operatorname{Tr}(\tau^a M)
$(SM~\cite{SM}).
Under a flavor rotation $U\in\mathrm{SU}(2)$, the matrix
$\Delta=\Delta^a\tau^a$ transforms as
$\Delta\rightarrow U\Delta U^\dagger$; Consequently, its three components
transform as an $\mathrm{SO}(3)$ vector.

At this stage $\bm\Delta=\bm u+i\bm v$ is a general complex triplet with six
real components. The uniform fermionic determinant depends on the two
invariants
$\varrho=\bm\Delta^*\cdot\bm\Delta$ and
$c=|i\bm\Delta^*\times\bm\Delta|$.
The two squared mass singular values are $m_\pm^2=\varrho\pm c$. At fixed
$\varrho$, their contribution to the effective potential satisfies (SM~\cite{SM})
\begin{align}
&V_{\rm eff}(\varrho,c)-V_{\rm eff}(\varrho,0)
\nonumber\\
&\qquad=
-2\int\frac{d^4k}{(2\pi)^4}
\ln\left[
1-\frac{c^2}{(k^2+\varrho)^2}
\right]
\geq 0.
\label{eq:collinear}
\end{align}
The minimum therefore occurs at $c=0$, which requires $\bm u$ and $\bm v$ to
be parallel. Fermionic energetics thus locks the general complex triplet into
the collinear form
\begin{equation}
\Delta^a=\Phi n^a e^{i\theta},
\qquad
\Phi\geq 0,
\qquad
\mathbf n^2=1.
\label{eq:order}
\end{equation}
Above the finite critical coupling, the gap equation fixes the amplitude at a
nonzero value $\Phi=\Phi_0$~(SM~\cite{SM}). 
Within the flavor-symmetric continuum model, the unit vector
$\mathbf n$ can point anywhere on $S^2$, while the density-wave phase
$\theta$ lies on $S^1$. However, these variables do not label distinct
states uniquely: for
$\Delta=\Delta_0 e^{i\theta}\mathbf n\cdot\boldsymbol{\tau}$, the simultaneous
change
$(\mathbf n,\theta)\to(-\mathbf n,\theta+\pi)$ leaves $\bm \Delta$ unchanged.
Thus, these two points must be identified, giving the order-parameter
manifold
\begin{equation}
\mathcal M
=
\frac{S^2\times S^1}{\mathbb Z_2}.
\label{eq:manifold}
\end{equation}

By contrast, the scalar Weyl CDW has the fixed-amplitude manifold $S^1$.
Flavor therefore introduces an internal orientational sector and ties it
nontrivially to the density-wave phase. This is shown schematically in Fig.~\ref{fig:1}.

\begin{figure*}
        \centering
        \includegraphics[width=1\textwidth]{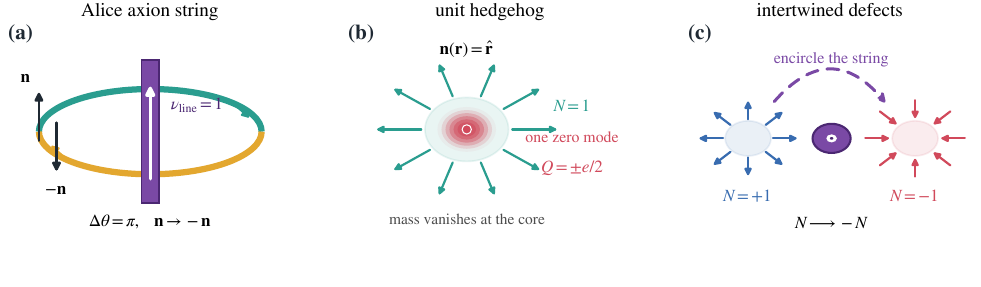}
       \caption{Topological defects and their fermionic consequences in the matrix density wave. (a) The elementary line defect is a half-quantum vortex in which the density-wave phase winds by $\pi$ while the flavor orientation reverses, $\mathbf n\rightarrow-\mathbf n$. Its mass determinant winds once, so the resulting Alice axion string carries a single chiral mode, $\nu_{\rm line}=1$. (b) A unit hedgehog,
$\mathbf n(\mathbf r)=\hat{\mathbf r}$, forces the adjoint mass to vanish at
its core and binds one normalizable zero mode with charge sectors
$Q=\pm e/2$ at neutrality in the spectrally symmetric limit.
(c) The line and point defects are intertwined: transporting a hedgehog around
the Alice string reverses its winding, $N\rightarrow-N$.} 
        \label{fig:2}
    \end{figure*}

\textit{Topological defects of the matrix density wave.}---
The enlarged order-parameter manifold enriches the defect structure in two
distinct ways: it modifies the elementary line defect inherited from the
scalar Weyl CDW and introduces a class of topologically stable point defects
absent in the scalar case. We focus here on the elementary representatives of
these two sectors---a half-quantum vortex and a unit hedgehog---and first
describe their topology and mutual relation.

Line defects are classified by $\pi_1(\mathcal M)=\mathbb Z$. Because
$(\mathbf n,\theta)\sim(-\mathbf n,\theta+\pi)$, its generator combines a
$\pi$ winding of the density-wave phase ($\theta$) with a reversal of the flavor
orientation. On a loop parametrized by the azimuthal angle $\varphi$, a
representative has $\theta(\varphi)=\theta_0+\varphi/2$, while
$\mathbf n(\varphi)$ rotates continuously from $\mathbf n_0$ to
$-\mathbf n_0$. Although neither component returns individually as
$\varphi$ advances from $0$ to $2\pi$, the full matrix order parameter
$\Delta(\varphi)=\Phi_0e^{i\theta(\varphi)}
\mathbf n(\varphi)\cdot\bm\tau$ does. This half-quantum vortex is therefore
the elementary line defect, whereas a $2\pi$ phase winding at fixed
$\mathbf n$ represents twice the generator.

Point defects are classified by maps from a sphere $S^2_\infty$ surrounding
the core into $\mathcal M$. Since $\pi_2(\mathcal M)=\mathbb Z$ and the
$S^1$ sector has trivial $\pi_2$, each class has a representative with a
constant density-wave phase. The nontrivial texture then resides in the flavor
orientation,
$\mathbf n:S^2_\infty\rightarrow S^2_{\mathrm{flavor}}$,
and is characterized by the integer winding number
\begin{equation}
N=
\frac{1}{4\pi}
\int_0^\pi d\vartheta
\int_0^{2\pi}d\varphi\,
\mathbf n\cdot
\left(
\partial_\vartheta\mathbf n
\times
\partial_\varphi\mathbf n
\right).
\label{eq:winding}
\end{equation}
Writing the real adjoint mass as
$\bm\Phi(\mathbf r)=\Phi_0f(\mathbf r)\mathbf n(\mathbf r)$, with
$f\rightarrow1$ at large distances, a texture with $N\neq0$ cannot remain
within the fixed-amplitude manifold throughout the enclosed volume;
otherwise, its nonzero-degree boundary map would extend continuously into
the ball. In the absence of a competing core order, the amplitude must
therefore vanish somewhere inside the enclosing sphere. The simplest
representative is the unit hedgehog,
\begin{equation}
\mathbf n(\mathbf r)=\hat{\mathbf r},
\qquad
\bm\Phi(\mathbf r)=\Phi_0f(r)\hat{\mathbf r},
\qquad
f(0)=0,\quad f(\infty)=1,
\label{eq:hedgehog}
\end{equation}
for which $N=1$. Since $\bm\Phi$ is the electronic mass responsible for the
bulk density-wave gap, its vanishing forces the gap to close at the defect
core.

Finally, the generator of $\pi_1(\mathcal M)$ acts nontrivially on the
point-defect winding number. It sends $\mathbf n\rightarrow-\mathbf n$, and
the antipodal map on $S^2$ has degree $-1$; hence $N\rightarrow-N$.
Transporting a hedgehog around the half-quantum vortex therefore converts it
into an antihedgehog. This reversal of the point-defect winding upon encircling the half-quantum vortex is the defining property of an Alice string~\cite{bensonimbo}.
 The matrix density wave thus possesses an intertwined
point--line defect structure absent from the scalar Weyl CDW. These elementary defects, along with their fermionic
consequences stated next, are summarized in Fig.~\ref{fig:2}.


\textit{Fermionic states of the defects.}---
We first consider a straight Alice string. Let $\Delta(\varphi)$ denote the
asymptotic internode mass matrix on a loop transverse to the string. The net
chirality of the modes propagating along the defect is fixed by the index of
the transverse Dirac operator,
\begin{equation}
\nu_{\mathrm{line}}
=
N_{\mathrm{R}}-N_{\mathrm{L}}
=
\operatorname{wind}\!\left[\det\Delta\right].
\label{eq:line-index}
\end{equation}
where $N_{\mathrm{R}}$ and $N_{\mathrm{L}}$ are the numbers of right- and
left-moving modes, respectively ~\cite{jackiwrossi,weinberg1981}. For the adjoint mass,
$
\det\Delta=-\Phi_0^2e^{2i\theta}
$ (SM~\cite{SM})
Although the density-wave phase of the elementary Alice string winds only by
$\pi$, the phase of $\det\Delta$ winds by $2\pi$. Hence
$\nu_{\mathrm{line}}=1$: the Alice string carries a single chiral electronic
mode and is therefore an elementary Alice axion string. By comparison, a
$2\pi$ winding of the density-wave phase at fixed flavor orientation has
$\nu_{\mathrm{line}}=2$ and carries two chiral modes. The scalar CDW of a
single Weyl pair also has $\nu_{\mathrm{line}}=1$, but its elementary defect
requires a full $2\pi$ phase winding. Flavor thus converts the conventional
axion string into a half-quantum defect without reducing its electronic index.

We next consider the hedgehog. After absorbing the rapid internode modulation
by a chiral rotation and choosing the constant density-wave phase to be zero,
the fermions are described by
\begin{equation}
H
=
-i v_F\rho_z\bm\sigma\cdot\bm\nabla
+
\rho_x\bm\Phi(\mathbf r)\cdot\bm\tau.
\label{eq:H}
\end{equation}
The Hamiltonian possesses the spectral chiral symmetry
$
\{H,\rho_y\}=0,
$
which pairs every nonzero-energy eigenstate at energy $E$ with one at energy
$-E$. In the basis that diagonalizes $\rho_y$, the Hamiltonian is off
diagonal,
$
H=
\begin{pmatrix}
0&D^\dagger\\
D&0
\end{pmatrix}.
$
The Callias index is determined entirely by the normalized asymptotic mass
$\hat{\Phi}^a=\Phi^a/|\bm\Phi|$:
\begin{equation}
\operatorname{ind}D
=
\dim\ker D-\dim\ker D^\dagger
=N.
\label{eq:index}
\end{equation}
A unit hedgehog therefore requires one unpaired zero mode. More generally, the
winding fixes the imbalance between zero modes of opposite $\rho_y$
eigenvalue, independently of the detailed core
profile~\cite{callias,bottseeley,weinberg,SM}.

For the spherically symmetric texture in Eq.~\eqref{eq:hedgehog}, the
normalizable zero mode can be obtained explicitly. Let $\ket{S}$ denote the
pseudospin--flavor singlet satisfying
$
\left(
\bm\sigma\otimes\mathbb I_\tau
+
\mathbb I_\sigma\otimes\bm\tau
\right)\ket{S}=0.
$
The unit hedgehog binds the state
\begin{equation}
\Psi_0(\mathbf r)
=
\mathcal N
\exp\left[
-\frac{\Phi_0}{v_F}
\int_0^r dr'\,f(r')
\right]
\ket{\rho_y=+1}\otimes\ket{S}.
\label{eq:zeromode}
\end{equation}
The state is regular at the origin and decays with localization length
$\xi_0=v_F/\Phi_0$. An analysis of the remaining angular channels shows that
they contain no additional normalizable zero modes(SM~\cite{SM}). The two flavor
components therefore do not produce two independent defect states; instead,
flavor and Weyl pseudospin lock into a single electronic mode.

At neutrality, all negative-energy states are filled and all positive-energy
states are empty. The contributions of the paired nonzero-energy states cancel
relative to the defect-free neutral state because of the spectral symmetry.
The defect charge is therefore determined by the occupation of the unpaired
zero mode,
\begin{equation}
Q
=
e\left(
a_0^\dagger a_0-\frac{1}{2}
\right).
\label{eq:charge}
\end{equation}
The empty and occupied zero-mode sectors differ by one electron and carry
charges $-e/2$ and $+e/2$, respectively, in the spectrally symmetric limit,
as in the Jackiw--Rebbi mechanism~\cite{jackiwrebbi,goldstonewilczek}.

In a finite system with periodic boundary conditions, the total hedgehog
winding must vanish, so the natural configuration is a
hedgehog--antihedgehog pair. For a separation $d\gg\xi_0$, the two defects
support local modes of opposite eigenvalues of $\rho_y$. Their hybridization produces
a pair of states at energies $\pm E_0$, with
$E_0\sim e^{-d/\xi_0}$ (SM~\cite{SM}). At neutrality the negative-energy state is occupied
and the positive-energy state is empty. Occupying the latter adds one fermion;
for a symmetric, well-separated pair, its density is shared equally between
the two cores, and the excess charge accumulated near either core approaches
$e/2$. The pair therefore produces two complementary local signatures:
tunneling spectroscopy should reveal core-localized in-gap resonances whose
splitting decreases exponentially with defect separation, while a
charge-sensitive probe should detect their occupation-dependent fractional
charge. It thus provides a globally consistent finite-system realization of
the local fractional charge associated with an isolated hedgehog.



\textit{Discussion.}---
Closely related quotient manifolds and half defects occur in other settings, such as collinear spin-density waves and polar spin-$1$ condensates
~\cite{zhangdemlersachdev,kirovabrazovskii,zhou,leonhardtvolovik}, while Alice
point--line topology appears in relativistic models, spinor condensates, and
chiral superconductors~\cite{bensonimbo,ruostekoskianglin,volovik,blinova}. Defect-bound zero modes and fractional fermion number have been discussed for
Dirac-mass hedgehogs in three dimensions~\cite{hosur,herbutlu,shao} and for
Kekul\'e vortices in two-dimensional Dirac systems
~\cite{houchamon,chamon}. The matrix density wave brings these mechanisms together in a single charge-conserving, interaction-selected, translation-breaking electronic mass:
a fractionally charged hedgehog and a chiral Alice axion string arise as
defects of the same condensate, and transporting the former around the latter
converts it into an antihedgehog. To our knowledge, this is the first
charge-conserving electronic ordered phase to realize this complete structure.

Our mean-field analysis establishes the preference for the adjoint over the
singlet component within the Weyl-pseudospin-scalar internode sector for the
specified local interaction; competition with other particle--hole and
particle--particle orders lies outside this restricted theory. Once the matrix
density wave is established, its defect topology and fermionic consequences
have distinct protection conditions. The hedgehog winding is protected by
$\pi_2(\mathcal M)=\mathbb Z$ as long as the order-parameter manifold is
preserved. For fixed winding, the Callias index protects an unpaired zero mode
under smooth deformations that preserve the spectral chiral symmetry and the
asymptotic mass gap. Exact half-charge quantization additionally requires
neutrality and a definite zero-mode occupation. Perturbations that break the
spectral chiral symmetry can shift the state away from zero and make its charge
nonuniversal, even when a localized in-gap state survives(SM~\cite{SM}). In a well-separated
hedgehog--antihedgehog pair, these states should appear as core-localized
resonances with exponentially small splitting, accompanied by an
occupation-dependent charge approaching $e/2$ at each core.

Beyond the elementary Alice string and unit hedgehog, $\mathcal M$ also permits
Hopf textures and composite configurations of hedgehogs and closed Alice-string
loops~\cite{kobayashi}, whose fermionic consequences remain open. Additionally, it remains to be seen what happens when the exact flavor $\mathrm{SU}(2)$ and continuous sliding symmetries of our
minimal model are relaxed. Flavor-mixing terms and anisotropic interactions can pin
$\mathbf n$, changing the order-parameter manifold and removing strict
hedgehog protection, while lattice commensurability can pin $\theta$ and
confine Alice strings through attached domain walls. Determining which features
survive in higher-charge and multifold semimetals will require incorporating
their microscopic symmetries and interactions.


In summary, we have analyzed interaction-driven internode density-wave order
in a two-flavor Weyl system. Within this sector, a local flavor-symmetric
repulsion favors a traceless matrix mass with manifold
$\mathcal M=(S^2\times S^1)/\mathbb Z_2$. This single condensate supports a
fractionally charged hedgehog and a chiral Alice axion string, and transporting
the hedgehog around the string converts it into an antihedgehog. Flavor thereby
provides a unified, interaction-driven realization of intertwined line and
point defects in a three-dimensional, charge-conserving electronic system.

\begin{acknowledgments}
S.M. acknowledges financial support from the Anusandhan National Research Foundation
(ANRF), India through the National Postdoctoral Fellowship (PDF/2025/003070). We acknowledge assistance from AI platforms ChatGPT and Claude.
\end{acknowledgments}

\bibliography{2weylcdw_references}

\begin{thebibliography}{36}%
\makeatletter
\providecommand \@ifxundefined [1]{%
 \@ifx{#1\undefined}
}%
\providecommand \@ifnum [1]{%
 \ifnum #1\expandafter \@firstoftwo
 \else \expandafter \@secondoftwo
 \fi
}%
\providecommand \@ifx [1]{%
 \ifx #1\expandafter \@firstoftwo
 \else \expandafter \@secondoftwo
 \fi
}%
\providecommand \natexlab [1]{#1}%
\providecommand \enquote  [1]{``#1''}%
\providecommand \bibnamefont  [1]{#1}%
\providecommand \bibfnamefont [1]{#1}%
\providecommand \citenamefont [1]{#1}%
\providecommand \href@noop [0]{\@secondoftwo}%
\providecommand \href [0]{\begingroup \@sanitize@url \@href}%
\providecommand \@href[1]{\@@startlink{#1}\@@href}%
\providecommand \@@href[1]{\endgroup#1\@@endlink}%
\providecommand \@sanitize@url [0]{\catcode `\\12\catcode `\$12\catcode
  `\&12\catcode `\#12\catcode `\^12\catcode `\_12\catcode `\%12\relax}%
\providecommand \@@startlink[1]{}%
\providecommand \@@endlink[0]{}%
\providecommand \url  [0]{\begingroup\@sanitize@url \@url }%
\providecommand \@url [1]{\endgroup\@href {#1}{\urlprefix }}%
\providecommand \urlprefix  [0]{URL }%
\providecommand \Eprint [0]{\href }%
\providecommand \doibase [0]{https://doi.org/}%
\providecommand \selectlanguage [0]{\@gobble}%
\providecommand \bibinfo  [0]{\@secondoftwo}%
\providecommand \bibfield  [0]{\@secondoftwo}%
\providecommand \translation [1]{[#1]}%
\providecommand \BibitemOpen [0]{}%
\providecommand \bibitemStop [0]{}%
\providecommand \bibitemNoStop [0]{.\EOS\space}%
\providecommand \EOS [0]{\spacefactor3000\relax}%
\providecommand \BibitemShut  [1]{\csname bibitem#1\endcsname}%
\let\auto@bib@innerbib\@empty
\bibitem [{\citenamefont {Wei}\ \emph {et~al.}(2012)\citenamefont {Wei},
  \citenamefont {Chao},\ and\ \citenamefont {Aji}}]{weiaji}%
  \BibitemOpen
  \bibfield  {author} {\bibinfo {author} {\bibfnamefont {H.}~\bibnamefont
  {Wei}}, \bibinfo {author} {\bibfnamefont {S.-P.}\ \bibnamefont {Chao}},\ and\
  \bibinfo {author} {\bibfnamefont {V.}~\bibnamefont {Aji}},\ }\bibfield
  {title} {\bibinfo {title} {Excitonic phases from {Weyl} semimetals},\ }\href
  {https://doi.org/10.1103/PhysRevLett.109.196403} {\bibfield  {journal}
  {\bibinfo  {journal} {Phys. Rev. Lett.}\ }\textbf {\bibinfo {volume} {109}},\
  \bibinfo {pages} {196403} (\bibinfo {year} {2012})}\BibitemShut {NoStop}%
\bibitem [{\citenamefont {Wang}\ and\ \citenamefont {Zhang}(2013)}]{wangzhang}%
  \BibitemOpen
  \bibfield  {author} {\bibinfo {author} {\bibfnamefont {Z.}~\bibnamefont
  {Wang}}\ and\ \bibinfo {author} {\bibfnamefont {S.-C.}\ \bibnamefont
  {Zhang}},\ }\bibfield  {title} {\bibinfo {title} {Chiral anomaly, charge
  density waves, and axion strings from {Weyl} semimetals},\ }\href
  {https://doi.org/10.1103/PhysRevB.87.161107} {\bibfield  {journal} {\bibinfo
  {journal} {Phys. Rev. B}\ }\textbf {\bibinfo {volume} {87}},\ \bibinfo
  {pages} {161107(R)} (\bibinfo {year} {2013})}\BibitemShut {NoStop}%
\bibitem [{\citenamefont {Roy}\ and\ \citenamefont {Sau}(2015)}]{roysau}%
  \BibitemOpen
  \bibfield  {author} {\bibinfo {author} {\bibfnamefont {B.}~\bibnamefont
  {Roy}}\ and\ \bibinfo {author} {\bibfnamefont {J.~D.}\ \bibnamefont {Sau}},\
  }\bibfield  {title} {\bibinfo {title} {Magnetic catalysis and axionic charge
  density wave in {Weyl} semimetals},\ }\href
  {https://doi.org/10.1103/PhysRevB.92.125141} {\bibfield  {journal} {\bibinfo
  {journal} {Phys. Rev. B}\ }\textbf {\bibinfo {volume} {92}},\ \bibinfo
  {pages} {125141} (\bibinfo {year} {2015})}\BibitemShut {NoStop}%
\bibitem [{\citenamefont {You}\ \emph {et~al.}(2016)\citenamefont {You},
  \citenamefont {Cho},\ and\ \citenamefont {Hughes}}]{youcho}%
  \BibitemOpen
  \bibfield  {author} {\bibinfo {author} {\bibfnamefont {Y.}~\bibnamefont
  {You}}, \bibinfo {author} {\bibfnamefont {G.~Y.}\ \bibnamefont {Cho}},\ and\
  \bibinfo {author} {\bibfnamefont {T.~L.}\ \bibnamefont {Hughes}},\ }\bibfield
   {title} {\bibinfo {title} {Response properties of axion insulators and
  {Weyl} semimetals driven by screw dislocations and dynamical axion strings},\
  }\href {https://doi.org/10.1103/PhysRevB.94.085102} {\bibfield  {journal}
  {\bibinfo  {journal} {Phys. Rev. B}\ }\textbf {\bibinfo {volume} {94}},\
  \bibinfo {pages} {085102} (\bibinfo {year} {2016})}\BibitemShut {NoStop}%
\bibitem [{\citenamefont {Callan}\ and\ \citenamefont
  {Harvey}(1985)}]{callanharvey}%
  \BibitemOpen
  \bibfield  {author} {\bibinfo {author} {\bibfnamefont {C.~G.}\ \bibnamefont
  {Callan}, \bibfnamefont {Jr.}}\ and\ \bibinfo {author} {\bibfnamefont
  {J.~A.}\ \bibnamefont {Harvey}},\ }\bibfield  {title} {\bibinfo {title}
  {Anomalies and fermion zero modes on strings and domain walls},\ }\href
  {https://doi.org/10.1016/0550-3213(85)90489-4} {\bibfield  {journal}
  {\bibinfo  {journal} {Nucl. Phys. B}\ }\textbf {\bibinfo {volume} {250}},\
  \bibinfo {pages} {427} (\bibinfo {year} {1985})}\BibitemShut {NoStop}%
\bibitem [{\citenamefont {Liu}\ \emph {et~al.}(2022)\citenamefont {Liu},
  \citenamefont {Zhang}, \citenamefont {Han},\ and\ \citenamefont
  {Liu}}]{flavorweyl}%
  \BibitemOpen
  \bibfield  {author} {\bibinfo {author} {\bibfnamefont {P.}~\bibnamefont
  {Liu}}, \bibinfo {author} {\bibfnamefont {A.}~\bibnamefont {Zhang}}, \bibinfo
  {author} {\bibfnamefont {J.}~\bibnamefont {Han}},\ and\ \bibinfo {author}
  {\bibfnamefont {Q.}~\bibnamefont {Liu}},\ }\bibfield  {title} {\bibinfo
  {title} {Chiral {Dirac}-like fermion in spin-orbit-free antiferromagnetic
  semimetals},\ }\href {https://doi.org/10.1016/j.xinn.2022.100343} {\bibfield
  {journal} {\bibinfo  {journal} {The Innovation}\ }\textbf {\bibinfo {volume}
  {3}},\ \bibinfo {pages} {100343} (\bibinfo {year} {2022})},\ \Eprint
  {https://arxiv.org/abs/2107.09984} {arXiv:2107.09984} \BibitemShut {NoStop}%
\bibitem [{\citenamefont {Bradlyn}\ \emph {et~al.}(2016)\citenamefont
  {Bradlyn}, \citenamefont {Cano}, \citenamefont {Wang}, \citenamefont
  {Vergniory}, \citenamefont {Felser}, \citenamefont {Cava},\ and\
  \citenamefont {Bernevig}}]{multifoldtheory}%
  \BibitemOpen
  \bibfield  {author} {\bibinfo {author} {\bibfnamefont {B.}~\bibnamefont
  {Bradlyn}}, \bibinfo {author} {\bibfnamefont {J.}~\bibnamefont {Cano}},
  \bibinfo {author} {\bibfnamefont {Z.}~\bibnamefont {Wang}}, \bibinfo {author}
  {\bibfnamefont {M.~G.}\ \bibnamefont {Vergniory}}, \bibinfo {author}
  {\bibfnamefont {C.}~\bibnamefont {Felser}}, \bibinfo {author} {\bibfnamefont
  {R.~J.}\ \bibnamefont {Cava}},\ and\ \bibinfo {author} {\bibfnamefont
  {B.~A.}\ \bibnamefont {Bernevig}},\ }\bibfield  {title} {\bibinfo {title}
  {Beyond {Dirac} and {Weyl} fermions: Unconventional quasiparticles in
  conventional crystals},\ }\href {https://doi.org/10.1126/science.aaf5037}
  {\bibfield  {journal} {\bibinfo  {journal} {Science}\ }\textbf {\bibinfo
  {volume} {353}},\ \bibinfo {pages} {aaf5037} (\bibinfo {year}
  {2016})}\BibitemShut {NoStop}%
\bibitem [{\citenamefont {Rao}\ \emph {et~al.}(2019)\citenamefont {Rao},
  \citenamefont {Li}, \citenamefont {Zhang}, \citenamefont {Tian},
  \citenamefont {Li}, \citenamefont {Fu}, \citenamefont {Tang}, \citenamefont
  {Wang}, \citenamefont {Li}, \citenamefont {Fan}, \citenamefont {Li},
  \citenamefont {Huang}, \citenamefont {Liu}, \citenamefont {Long},
  \citenamefont {Fang}, \citenamefont {Weng}, \citenamefont {Shi},
  \citenamefont {Lei}, \citenamefont {Sun}, \citenamefont {Qian},\ and\
  \citenamefont {Ding}}]{multifoldexp}%
  \BibitemOpen
  \bibfield  {author} {\bibinfo {author} {\bibfnamefont {Z.}~\bibnamefont
  {Rao}}, \bibinfo {author} {\bibfnamefont {H.}~\bibnamefont {Li}}, \bibinfo
  {author} {\bibfnamefont {T.}~\bibnamefont {Zhang}}, \bibinfo {author}
  {\bibfnamefont {S.}~\bibnamefont {Tian}}, \bibinfo {author} {\bibfnamefont
  {C.}~\bibnamefont {Li}}, \bibinfo {author} {\bibfnamefont {B.}~\bibnamefont
  {Fu}}, \bibinfo {author} {\bibfnamefont {C.}~\bibnamefont {Tang}}, \bibinfo
  {author} {\bibfnamefont {L.}~\bibnamefont {Wang}}, \bibinfo {author}
  {\bibfnamefont {Z.}~\bibnamefont {Li}}, \bibinfo {author} {\bibfnamefont
  {W.}~\bibnamefont {Fan}}, \bibinfo {author} {\bibfnamefont {J.}~\bibnamefont
  {Li}}, \bibinfo {author} {\bibfnamefont {Y.}~\bibnamefont {Huang}}, \bibinfo
  {author} {\bibfnamefont {Z.}~\bibnamefont {Liu}}, \bibinfo {author}
  {\bibfnamefont {Y.}~\bibnamefont {Long}}, \bibinfo {author} {\bibfnamefont
  {C.}~\bibnamefont {Fang}}, \bibinfo {author} {\bibfnamefont {H.}~\bibnamefont
  {Weng}}, \bibinfo {author} {\bibfnamefont {Y.}~\bibnamefont {Shi}}, \bibinfo
  {author} {\bibfnamefont {H.}~\bibnamefont {Lei}}, \bibinfo {author}
  {\bibfnamefont {Y.}~\bibnamefont {Sun}}, \bibinfo {author} {\bibfnamefont
  {T.}~\bibnamefont {Qian}},\ and\ \bibinfo {author} {\bibfnamefont
  {H.}~\bibnamefont {Ding}},\ }\bibfield  {title} {\bibinfo {title}
  {Observation of unconventional chiral fermions with long {Fermi} arcs in
  {CoSi}},\ }\href {https://doi.org/10.1038/s41586-019-1031-8} {\bibfield
  {journal} {\bibinfo  {journal} {Nature}\ }\textbf {\bibinfo {volume} {567}},\
  \bibinfo {pages} {496} (\bibinfo {year} {2019})}\BibitemShut {NoStop}%
\bibitem [{\citenamefont {Sanchez}\ \emph {et~al.}(2019)\citenamefont
  {Sanchez}, \citenamefont {Belopolski}, \citenamefont {Cochran}, \citenamefont
  {Xu}, \citenamefont {Yin}, \citenamefont {Chang} \emph
  {et~al.}}]{sanchezrhsi}%
  \BibitemOpen
  \bibfield  {author} {\bibinfo {author} {\bibfnamefont {D.~S.}\ \bibnamefont
  {Sanchez}}, \bibinfo {author} {\bibfnamefont {I.}~\bibnamefont {Belopolski}},
  \bibinfo {author} {\bibfnamefont {T.~A.}\ \bibnamefont {Cochran}}, \bibinfo
  {author} {\bibfnamefont {X.}~\bibnamefont {Xu}}, \bibinfo {author}
  {\bibfnamefont {J.-X.}\ \bibnamefont {Yin}}, \bibinfo {author} {\bibfnamefont
  {G.}~\bibnamefont {Chang}}, \emph {et~al.},\ }\bibfield  {title} {\bibinfo
  {title} {Topological chiral crystals with helicoid-arc quantum states},\
  }\href {https://doi.org/10.1038/s41586-019-1037-2} {\bibfield  {journal}
  {\bibinfo  {journal} {Nature}\ }\textbf {\bibinfo {volume} {567}},\ \bibinfo
  {pages} {500} (\bibinfo {year} {2019})}\BibitemShut {NoStop}%
\bibitem [{\citenamefont {Schr{\"o}ter}\ \emph {et~al.}(2019)\citenamefont
  {Schr{\"o}ter}, \citenamefont {Pei}, \citenamefont {Vergniory}, \citenamefont
  {Sun}, \citenamefont {Manna}, \citenamefont {de~Juan} \emph
  {et~al.}}]{schroeteralpt}%
  \BibitemOpen
  \bibfield  {author} {\bibinfo {author} {\bibfnamefont {N.~B.~M.}\
  \bibnamefont {Schr{\"o}ter}}, \bibinfo {author} {\bibfnamefont
  {D.}~\bibnamefont {Pei}}, \bibinfo {author} {\bibfnamefont {M.~G.}\
  \bibnamefont {Vergniory}}, \bibinfo {author} {\bibfnamefont {Y.}~\bibnamefont
  {Sun}}, \bibinfo {author} {\bibfnamefont {K.}~\bibnamefont {Manna}}, \bibinfo
  {author} {\bibfnamefont {F.}~\bibnamefont {de~Juan}}, \emph {et~al.},\
  }\bibfield  {title} {\bibinfo {title} {Chiral topological semimetal with
  multifold band crossings and long {Fermi} arcs},\ }\href
  {https://doi.org/10.1038/s41567-019-0511-y} {\bibfield  {journal} {\bibinfo
  {journal} {Nature Physics}\ }\textbf {\bibinfo {volume} {15}},\ \bibinfo
  {pages} {759} (\bibinfo {year} {2019})}\BibitemShut {NoStop}%
\bibitem [{\citenamefont {Dantas}\ \emph {et~al.}(2020)\citenamefont {Dantas},
  \citenamefont {Pe{\~n}a-Benitez}, \citenamefont {Roy},\ and\ \citenamefont
  {Sur{\'o}wka}}]{multiflavorhigher}%
  \BibitemOpen
  \bibfield  {author} {\bibinfo {author} {\bibfnamefont {R.~M.~A.}\
  \bibnamefont {Dantas}}, \bibinfo {author} {\bibfnamefont {F.}~\bibnamefont
  {Pe{\~n}a-Benitez}}, \bibinfo {author} {\bibfnamefont {B.}~\bibnamefont
  {Roy}},\ and\ \bibinfo {author} {\bibfnamefont {P.}~\bibnamefont
  {Sur{\'o}wka}},\ }\bibfield  {title} {\bibinfo {title} {Non-{Abelian}
  anomalies in multi-{Weyl} semimetals},\ }\href
  {https://doi.org/10.1103/PhysRevResearch.2.013007} {\bibfield  {journal}
  {\bibinfo  {journal} {Phys. Rev. Research}\ }\textbf {\bibinfo {volume}
  {2}},\ \bibinfo {pages} {013007} (\bibinfo {year} {2020})},\ \Eprint
  {https://arxiv.org/abs/1905.02189} {arXiv:1905.02189} \BibitemShut {NoStop}%
\bibitem [{\citenamefont {Mukherjee}\ \emph {et~al.}(2026)\citenamefont
  {Mukherjee}, \citenamefont {Sharma},\ and\ \citenamefont {Pal}}]{mukherjee}%
  \BibitemOpen
  \bibfield  {author} {\bibinfo {author} {\bibfnamefont {S.}~\bibnamefont
  {Mukherjee}}, \bibinfo {author} {\bibfnamefont {S.}~\bibnamefont {Sharma}},\
  and\ \bibinfo {author} {\bibfnamefont {H.~K.}\ \bibnamefont {Pal}},\
  }\bibfield  {title} {\bibinfo {title} {Anomalous chiral anomaly in spin-1
  fermionic systems},\ }\href {https://doi.org/10.1103/tz3b-lfk5} {\bibfield
  {journal} {\bibinfo  {journal} {Phys. Rev. B}\ }\textbf {\bibinfo {volume}
  {113}},\ \bibinfo {pages} {075139} (\bibinfo {year} {2026})}\BibitemShut
  {NoStop}%
\bibitem [{\citenamefont {Wang}\ \emph {et~al.}(2021)\citenamefont {Wang},
  \citenamefont {Cheng}, \citenamefont {Wang}, \citenamefont {Zhang},
  \citenamefont {Lu}, \citenamefont {Yi}, \citenamefont {Niu}, \citenamefont
  {Deng}, \citenamefont {Liu}, \citenamefont {Chen},\ and\ \citenamefont
  {Pan}}]{coldatomweyl}%
  \BibitemOpen
  \bibfield  {author} {\bibinfo {author} {\bibfnamefont {Z.-Y.}\ \bibnamefont
  {Wang}}, \bibinfo {author} {\bibfnamefont {X.-C.}\ \bibnamefont {Cheng}},
  \bibinfo {author} {\bibfnamefont {B.-Z.}\ \bibnamefont {Wang}}, \bibinfo
  {author} {\bibfnamefont {J.-Y.}\ \bibnamefont {Zhang}}, \bibinfo {author}
  {\bibfnamefont {Y.-H.}\ \bibnamefont {Lu}}, \bibinfo {author} {\bibfnamefont
  {C.-R.}\ \bibnamefont {Yi}}, \bibinfo {author} {\bibfnamefont
  {S.}~\bibnamefont {Niu}}, \bibinfo {author} {\bibfnamefont {Y.}~\bibnamefont
  {Deng}}, \bibinfo {author} {\bibfnamefont {X.-J.}\ \bibnamefont {Liu}},
  \bibinfo {author} {\bibfnamefont {S.}~\bibnamefont {Chen}},\ and\ \bibinfo
  {author} {\bibfnamefont {J.-W.}\ \bibnamefont {Pan}},\ }\bibfield  {title}
  {\bibinfo {title} {Realization of an ideal {Weyl} semimetal band in a quantum
  gas with {3D} spin-orbit coupling},\ }\href
  {https://doi.org/10.1126/science.abc0105} {\bibfield  {journal} {\bibinfo
  {journal} {Science}\ }\textbf {\bibinfo {volume} {372}},\ \bibinfo {pages}
  {271} (\bibinfo {year} {2021})},\ \Eprint {https://arxiv.org/abs/2004.02413}
  {arXiv:2004.02413} \BibitemShut {NoStop}%
\bibitem [{Supplemental Material()}]{SM}%
  \BibitemOpen
  Supplemental Material,\ \href@noop {} {}\bibinfo {note} {See Supplemental
  Material for the Hartree--Fock channel decomposition and adjoint-channel
  selection; Hubbard--Stratonovich decoupling and the order-parameter manifold;
  the uniform adjoint-CDW effective potential, gap equation, and gradient
  expansion; hedgehog topology; the index of the Alice-string fermion mode;
  monopole-bound fermion zero mode, index theorem, and higher-winding monopole
  textures; and charge fractionalization for isolated and paired
  defects.}\BibitemShut {Stop}%
\bibitem [{\citenamefont {Nambu}\ and\ \citenamefont
  {Jona-Lasinio}(1961)}]{njl}%
  \BibitemOpen
  \bibfield  {author} {\bibinfo {author} {\bibfnamefont {Y.}~\bibnamefont
  {Nambu}}\ and\ \bibinfo {author} {\bibfnamefont {G.}~\bibnamefont
  {Jona-Lasinio}},\ }\bibfield  {title} {\bibinfo {title} {Dynamical model of
  elementary particles based on an analogy with superconductivity. {I}},\
  }\href {https://doi.org/10.1103/PhysRev.122.345} {\bibfield  {journal}
  {\bibinfo  {journal} {Phys. Rev.}\ }\textbf {\bibinfo {volume} {122}},\
  \bibinfo {pages} {345} (\bibinfo {year} {1961})}\BibitemShut {NoStop}%
\bibitem [{\citenamefont {Benson}\ and\ \citenamefont
  {Imbo}(2004)}]{bensonimbo}%
  \BibitemOpen
  \bibfield  {author} {\bibinfo {author} {\bibfnamefont {K.~M.}\ \bibnamefont
  {Benson}}\ and\ \bibinfo {author} {\bibfnamefont {T.}~\bibnamefont {Imbo}},\
  }\bibfield  {title} {\bibinfo {title} {Topologically {Alice} strings and
  monopoles},\ }\href {https://doi.org/10.1103/PhysRevD.70.025005} {\bibfield
  {journal} {\bibinfo  {journal} {Phys. Rev. D}\ }\textbf {\bibinfo {volume}
  {70}},\ \bibinfo {pages} {025005} (\bibinfo {year} {2004})},\ \Eprint
  {https://arxiv.org/abs/hep-th/0407001} {arXiv:hep-th/0407001} \BibitemShut
  {NoStop}%
\bibitem [{\citenamefont {Jackiw}\ and\ \citenamefont
  {Rossi}(1981)}]{jackiwrossi}%
  \BibitemOpen
  \bibfield  {author} {\bibinfo {author} {\bibfnamefont {R.}~\bibnamefont
  {Jackiw}}\ and\ \bibinfo {author} {\bibfnamefont {P.}~\bibnamefont {Rossi}},\
  }\bibfield  {title} {\bibinfo {title} {Zero modes of the vortex--fermion
  system},\ }\href {https://doi.org/10.1016/0550-3213(81)90044-4} {\bibfield
  {journal} {\bibinfo  {journal} {Nucl. Phys. B}\ }\textbf {\bibinfo {volume}
  {190}},\ \bibinfo {pages} {681} (\bibinfo {year} {1981})}\BibitemShut
  {NoStop}%
\bibitem [{\citenamefont {Weinberg}(1981)}]{weinberg1981}%
  \BibitemOpen
  \bibfield  {author} {\bibinfo {author} {\bibfnamefont {E.~J.}\ \bibnamefont
  {Weinberg}},\ }\bibfield  {title} {\bibinfo {title} {Index calculations for
  the fermion--vortex system},\ }\href
  {https://doi.org/10.1103/PhysRevD.24.2669} {\bibfield  {journal} {\bibinfo
  {journal} {Phys. Rev. D}\ }\textbf {\bibinfo {volume} {24}},\ \bibinfo
  {pages} {2669} (\bibinfo {year} {1981})}\BibitemShut {NoStop}%
\bibitem [{\citenamefont {Callias}(1978)}]{callias}%
  \BibitemOpen
  \bibfield  {author} {\bibinfo {author} {\bibfnamefont {C.}~\bibnamefont
  {Callias}},\ }\bibfield  {title} {\bibinfo {title} {Axial anomalies and index
  theorems on open spaces},\ }\href {https://doi.org/10.1007/BF01202525}
  {\bibfield  {journal} {\bibinfo  {journal} {Commun. Math. Phys.}\ }\textbf
  {\bibinfo {volume} {62}},\ \bibinfo {pages} {213} (\bibinfo {year}
  {1978})}\BibitemShut {NoStop}%
\bibitem [{\citenamefont {Bott}\ and\ \citenamefont
  {Seeley}(1978)}]{bottseeley}%
  \BibitemOpen
  \bibfield  {author} {\bibinfo {author} {\bibfnamefont {R.}~\bibnamefont
  {Bott}}\ and\ \bibinfo {author} {\bibfnamefont {R.}~\bibnamefont {Seeley}},\
  }\bibfield  {title} {\bibinfo {title} {Some remarks on the paper of
  {Callias}},\ }\href {https://doi.org/10.1007/BF01202526} {\bibfield
  {journal} {\bibinfo  {journal} {Commun. Math. Phys.}\ }\textbf {\bibinfo
  {volume} {62}},\ \bibinfo {pages} {235} (\bibinfo {year} {1978})}\BibitemShut
  {NoStop}%
\bibitem [{\citenamefont {Weinberg}(1979)}]{weinberg}%
  \BibitemOpen
  \bibfield  {author} {\bibinfo {author} {\bibfnamefont {E.~J.}\ \bibnamefont
  {Weinberg}},\ }\bibfield  {title} {\bibinfo {title} {Parameter counting for
  multimonopole solutions},\ }\href {https://doi.org/10.1103/PhysRevD.20.936}
  {\bibfield  {journal} {\bibinfo  {journal} {Phys. Rev. D}\ }\textbf {\bibinfo
  {volume} {20}},\ \bibinfo {pages} {936} (\bibinfo {year} {1979})}\BibitemShut
  {NoStop}%
\bibitem [{\citenamefont {Jackiw}\ and\ \citenamefont
  {Rebbi}(1976)}]{jackiwrebbi}%
  \BibitemOpen
  \bibfield  {author} {\bibinfo {author} {\bibfnamefont {R.}~\bibnamefont
  {Jackiw}}\ and\ \bibinfo {author} {\bibfnamefont {C.}~\bibnamefont {Rebbi}},\
  }\bibfield  {title} {\bibinfo {title} {Solitons with fermion number 1/2},\
  }\href {https://doi.org/10.1103/PhysRevD.13.3398} {\bibfield  {journal}
  {\bibinfo  {journal} {Phys. Rev. D}\ }\textbf {\bibinfo {volume} {13}},\
  \bibinfo {pages} {3398} (\bibinfo {year} {1976})}\BibitemShut {NoStop}%
\bibitem [{\citenamefont {Goldstone}\ and\ \citenamefont
  {Wilczek}(1981)}]{goldstonewilczek}%
  \BibitemOpen
  \bibfield  {author} {\bibinfo {author} {\bibfnamefont {J.}~\bibnamefont
  {Goldstone}}\ and\ \bibinfo {author} {\bibfnamefont {F.}~\bibnamefont
  {Wilczek}},\ }\bibfield  {title} {\bibinfo {title} {Fractional quantum
  numbers on solitons},\ }\href {https://doi.org/10.1103/PhysRevLett.47.986}
  {\bibfield  {journal} {\bibinfo  {journal} {Phys. Rev. Lett.}\ }\textbf
  {\bibinfo {volume} {47}},\ \bibinfo {pages} {986} (\bibinfo {year}
  {1981})}\BibitemShut {NoStop}%
\bibitem [{\citenamefont {Zhang}\ \emph {et~al.}(2002)\citenamefont {Zhang},
  \citenamefont {Demler},\ and\ \citenamefont {Sachdev}}]{zhangdemlersachdev}%
  \BibitemOpen
  \bibfield  {author} {\bibinfo {author} {\bibfnamefont {Y.}~\bibnamefont
  {Zhang}}, \bibinfo {author} {\bibfnamefont {E.}~\bibnamefont {Demler}},\ and\
  \bibinfo {author} {\bibfnamefont {S.}~\bibnamefont {Sachdev}},\ }\bibfield
  {title} {\bibinfo {title} {Competing orders in a magnetic field: Spin and
  charge order in the cuprate superconductors},\ }\href
  {https://doi.org/10.1103/PhysRevB.66.094501} {\bibfield  {journal} {\bibinfo
  {journal} {Phys. Rev. B}\ }\textbf {\bibinfo {volume} {66}},\ \bibinfo
  {pages} {094501} (\bibinfo {year} {2002})},\ \Eprint
  {https://arxiv.org/abs/cond-mat/0112343} {arXiv:cond-mat/0112343}
  \BibitemShut {NoStop}%
\bibitem [{\citenamefont {Kirova}\ and\ \citenamefont
  {Brazovskii}(2000)}]{kirovabrazovskii}%
  \BibitemOpen
  \bibfield  {author} {\bibinfo {author} {\bibfnamefont {N.}~\bibnamefont
  {Kirova}}\ and\ \bibinfo {author} {\bibfnamefont {S.}~\bibnamefont
  {Brazovskii}},\ }\bibfield  {title} {\bibinfo {title} {Topological defects in
  spin density waves},\ }\href {https://doi.org/10.1051/jp4:2000320} {\bibfield
   {journal} {\bibinfo  {journal} {J. Phys. IV France}\ }\textbf {\bibinfo
  {volume} {10}},\ \bibinfo {pages} {183} (\bibinfo {year} {2000})},\ \Eprint
  {https://arxiv.org/abs/cond-mat/0004313} {arXiv:cond-mat/0004313}
  \BibitemShut {NoStop}%
\bibitem [{\citenamefont {Zhou}(2001)}]{zhou}%
  \BibitemOpen
  \bibfield  {author} {\bibinfo {author} {\bibfnamefont {F.}~\bibnamefont
  {Zhou}},\ }\bibfield  {title} {\bibinfo {title} {Spin correlation and
  discrete symmetry in spinor {Bose--Einstein} condensates},\ }\href
  {https://doi.org/10.1103/PhysRevLett.87.080401} {\bibfield  {journal}
  {\bibinfo  {journal} {Phys. Rev. Lett.}\ }\textbf {\bibinfo {volume} {87}},\
  \bibinfo {pages} {080401} (\bibinfo {year} {2001})},\ \Eprint
  {https://arxiv.org/abs/cond-mat/0102372} {arXiv:cond-mat/0102372}
  \BibitemShut {NoStop}%
\bibitem [{\citenamefont {Leonhardt}\ and\ \citenamefont
  {Volovik}(2000)}]{leonhardtvolovik}%
  \BibitemOpen
  \bibfield  {author} {\bibinfo {author} {\bibfnamefont {U.}~\bibnamefont
  {Leonhardt}}\ and\ \bibinfo {author} {\bibfnamefont {G.~E.}\ \bibnamefont
  {Volovik}},\ }\bibfield  {title} {\bibinfo {title} {How to create an {Alice}
  string (half-quantum vortex) in a vector {Bose--Einstein} condensate},\
  }\href {https://doi.org/10.1134/1.1312008} {\bibfield  {journal} {\bibinfo
  {journal} {JETP Lett.}\ }\textbf {\bibinfo {volume} {72}},\ \bibinfo {pages}
  {46} (\bibinfo {year} {2000})},\ \Eprint
  {https://arxiv.org/abs/cond-mat/0003428} {arXiv:cond-mat/0003428}
  \BibitemShut {NoStop}%
\bibitem [{\citenamefont {Ruostekoski}\ and\ \citenamefont
  {Anglin}(2003)}]{ruostekoskianglin}%
  \BibitemOpen
  \bibfield  {author} {\bibinfo {author} {\bibfnamefont {J.}~\bibnamefont
  {Ruostekoski}}\ and\ \bibinfo {author} {\bibfnamefont {J.~R.}\ \bibnamefont
  {Anglin}},\ }\bibfield  {title} {\bibinfo {title} {Monopole core instability
  and {Alice} rings in spinor {Bose--Einstein} condensates},\ }\href
  {https://doi.org/10.1103/PhysRevLett.91.190402} {\bibfield  {journal}
  {\bibinfo  {journal} {Phys. Rev. Lett.}\ }\textbf {\bibinfo {volume} {91}},\
  \bibinfo {pages} {190402} (\bibinfo {year} {2003})},\ \bibinfo {note}
  {erratum: Phys. Rev. Lett. 97, 069902 (2006)},\ \Eprint
  {https://arxiv.org/abs/cond-mat/0307651} {arXiv:cond-mat/0307651}
  \BibitemShut {NoStop}%
\bibitem [{\citenamefont {Volovik}(2000)}]{volovik}%
  \BibitemOpen
  \bibfield  {author} {\bibinfo {author} {\bibfnamefont {G.~E.}\ \bibnamefont
  {Volovik}},\ }\bibfield  {title} {\bibinfo {title} {Monopoles and fractional
  vortices in chiral superconductors},\ }\href
  {https://doi.org/10.1073/pnas.97.6.2431} {\bibfield  {journal} {\bibinfo
  {journal} {Proc. Natl. Acad. Sci. USA}\ }\textbf {\bibinfo {volume} {97}},\
  \bibinfo {pages} {2431} (\bibinfo {year} {2000})},\ \Eprint
  {https://arxiv.org/abs/cond-mat/9911486} {arXiv:cond-mat/9911486}
  \BibitemShut {NoStop}%
\bibitem [{\citenamefont {Blinova}\ \emph {et~al.}(2023)\citenamefont
  {Blinova}, \citenamefont {Zamora-Zamora}, \citenamefont {Ollikainen},
  \citenamefont {Kivioja}, \citenamefont {M{\"o}tt{\"o}nen},\ and\
  \citenamefont {Hall}}]{blinova}%
  \BibitemOpen
  \bibfield  {author} {\bibinfo {author} {\bibfnamefont {A.}~\bibnamefont
  {Blinova}}, \bibinfo {author} {\bibfnamefont {R.}~\bibnamefont
  {Zamora-Zamora}}, \bibinfo {author} {\bibfnamefont {T.}~\bibnamefont
  {Ollikainen}}, \bibinfo {author} {\bibfnamefont {M.}~\bibnamefont {Kivioja}},
  \bibinfo {author} {\bibfnamefont {M.}~\bibnamefont {M{\"o}tt{\"o}nen}},\ and\
  \bibinfo {author} {\bibfnamefont {D.~S.}\ \bibnamefont {Hall}},\ }\bibfield
  {title} {\bibinfo {title} {Observation of an {Alice} ring in a
  {Bose--Einstein} condensate},\ }\href
  {https://doi.org/10.1038/s41467-023-40710-2} {\bibfield  {journal} {\bibinfo
  {journal} {Nat. Commun.}\ }\textbf {\bibinfo {volume} {14}},\ \bibinfo
  {pages} {5100} (\bibinfo {year} {2023})}\BibitemShut {NoStop}%
\bibitem [{\citenamefont {Hosur}\ \emph {et~al.}(2010)\citenamefont {Hosur},
  \citenamefont {Ryu},\ and\ \citenamefont {Vishwanath}}]{hosur}%
  \BibitemOpen
  \bibfield  {author} {\bibinfo {author} {\bibfnamefont {P.}~\bibnamefont
  {Hosur}}, \bibinfo {author} {\bibfnamefont {S.}~\bibnamefont {Ryu}},\ and\
  \bibinfo {author} {\bibfnamefont {A.}~\bibnamefont {Vishwanath}},\ }\bibfield
   {title} {\bibinfo {title} {Chiral topological insulators, superconductors,
  and other competing orders in three dimensions},\ }\href
  {https://doi.org/10.1103/PhysRevB.81.045120} {\bibfield  {journal} {\bibinfo
  {journal} {Phys. Rev. B}\ }\textbf {\bibinfo {volume} {81}},\ \bibinfo
  {pages} {045120} (\bibinfo {year} {2010})},\ \Eprint
  {https://arxiv.org/abs/0908.2691} {arXiv:0908.2691} \BibitemShut {NoStop}%
\bibitem [{\citenamefont {Herbut}\ and\ \citenamefont {Lu}(2011)}]{herbutlu}%
  \BibitemOpen
  \bibfield  {author} {\bibinfo {author} {\bibfnamefont {I.~F.}\ \bibnamefont
  {Herbut}}\ and\ \bibinfo {author} {\bibfnamefont {C.-K.}\ \bibnamefont
  {Lu}},\ }\bibfield  {title} {\bibinfo {title} {Spectrum of the {Dirac}
  hamiltonian with the mass hedgehog in arbitrary dimension},\ }\href
  {https://doi.org/10.1103/PhysRevB.83.125412} {\bibfield  {journal} {\bibinfo
  {journal} {Phys. Rev. B}\ }\textbf {\bibinfo {volume} {83}},\ \bibinfo
  {pages} {125412} (\bibinfo {year} {2011})},\ \Eprint
  {https://arxiv.org/abs/1010.0728} {arXiv:1010.0728} \BibitemShut {NoStop}%
\bibitem [{\citenamefont {Shao}(2025)}]{shao}%
  \BibitemOpen
  \bibfield  {author} {\bibinfo {author} {\bibfnamefont {L.~B.}\ \bibnamefont
  {Shao}},\ }\bibfield  {title} {\bibinfo {title} {Fractional {Fermi} numbers
  in {Dirac} systems with topological point defects},\ }\href
  {https://doi.org/10.1103/bqlw-2jlx} {\bibfield  {journal} {\bibinfo
  {journal} {Phys. Rev. B}\ }\textbf {\bibinfo {volume} {112}},\ \bibinfo
  {pages} {064110} (\bibinfo {year} {2025})}\BibitemShut {NoStop}%
\bibitem [{\citenamefont {Hou}\ \emph {et~al.}(2007)\citenamefont {Hou},
  \citenamefont {Chamon},\ and\ \citenamefont {Mudry}}]{houchamon}%
  \BibitemOpen
  \bibfield  {author} {\bibinfo {author} {\bibfnamefont {C.-Y.}\ \bibnamefont
  {Hou}}, \bibinfo {author} {\bibfnamefont {C.}~\bibnamefont {Chamon}},\ and\
  \bibinfo {author} {\bibfnamefont {C.}~\bibnamefont {Mudry}},\ }\bibfield
  {title} {\bibinfo {title} {Electron fractionalization in two-dimensional
  graphenelike structures},\ }\href
  {https://doi.org/10.1103/PhysRevLett.98.186809} {\bibfield  {journal}
  {\bibinfo  {journal} {Phys. Rev. Lett.}\ }\textbf {\bibinfo {volume} {98}},\
  \bibinfo {pages} {186809} (\bibinfo {year} {2007})}\BibitemShut {NoStop}%
\bibitem [{\citenamefont {Chamon}\ \emph {et~al.}(2008)\citenamefont {Chamon},
  \citenamefont {Hou}, \citenamefont {Jackiw}, \citenamefont {Mudry},
  \citenamefont {Pi},\ and\ \citenamefont {Semenoff}}]{chamon}%
  \BibitemOpen
  \bibfield  {author} {\bibinfo {author} {\bibfnamefont {C.}~\bibnamefont
  {Chamon}}, \bibinfo {author} {\bibfnamefont {C.-Y.}\ \bibnamefont {Hou}},
  \bibinfo {author} {\bibfnamefont {R.}~\bibnamefont {Jackiw}}, \bibinfo
  {author} {\bibfnamefont {C.}~\bibnamefont {Mudry}}, \bibinfo {author}
  {\bibfnamefont {S.-Y.}\ \bibnamefont {Pi}},\ and\ \bibinfo {author}
  {\bibfnamefont {G.~W.}\ \bibnamefont {Semenoff}},\ }\bibfield  {title}
  {\bibinfo {title} {Electron fractionalization for two-dimensional {Dirac}
  fermions},\ }\href {https://doi.org/10.1103/PhysRevB.77.235431} {\bibfield
  {journal} {\bibinfo  {journal} {Phys. Rev. B}\ }\textbf {\bibinfo {volume}
  {77}},\ \bibinfo {pages} {235431} (\bibinfo {year} {2008})}\BibitemShut
  {NoStop}%
\bibitem [{\citenamefont {Kobayashi}\ \emph {et~al.}(2012)\citenamefont
  {Kobayashi}, \citenamefont {Kobayashi}, \citenamefont {Kawaguchi},
  \citenamefont {Nitta},\ and\ \citenamefont {Ueda}}]{kobayashi}%
  \BibitemOpen
  \bibfield  {author} {\bibinfo {author} {\bibfnamefont {S.}~\bibnamefont
  {Kobayashi}}, \bibinfo {author} {\bibfnamefont {M.}~\bibnamefont
  {Kobayashi}}, \bibinfo {author} {\bibfnamefont {Y.}~\bibnamefont
  {Kawaguchi}}, \bibinfo {author} {\bibfnamefont {M.}~\bibnamefont {Nitta}},\
  and\ \bibinfo {author} {\bibfnamefont {M.}~\bibnamefont {Ueda}},\ }\bibfield
  {title} {\bibinfo {title} {{Abe} homotopy classification of topological
  excitations under the topological influence of vortices},\ }\href
  {https://doi.org/10.1016/j.nuclphysb.2011.11.003} {\bibfield  {journal}
  {\bibinfo  {journal} {Nucl. Phys. B}\ }\textbf {\bibinfo {volume} {856}},\
  \bibinfo {pages} {577} (\bibinfo {year} {2012})},\ \Eprint
  {https://arxiv.org/abs/1110.1478} {arXiv:1110.1478} \BibitemShut {NoStop}%
\end{thebibliography}%


\begin{thebibliography}{0}%
\makeatletter
\providecommand \@ifxundefined [1]{%
 \@ifx{#1\undefined}
}%
\providecommand \@ifnum [1]{%
 \ifnum #1\expandafter \@firstoftwo
 \else \expandafter \@secondoftwo
 \fi
}%
\providecommand \@ifx [1]{%
 \ifx #1\expandafter \@firstoftwo
 \else \expandafter \@secondoftwo
 \fi
}%
\providecommand \natexlab [1]{#1}%
\providecommand \enquote  [1]{``#1''}%
\providecommand \bibnamefont  [1]{#1}%
\providecommand \bibfnamefont [1]{#1}%
\providecommand \citenamefont [1]{#1}%
\providecommand \href@noop [0]{\@secondoftwo}%
\providecommand \href [0]{\begingroup \@sanitize@url \@href}%
\providecommand \@href[1]{\@@startlink{#1}\@@href}%
\providecommand \@@href[1]{\endgroup#1\@@endlink}%
\providecommand \@sanitize@url [0]{\catcode `\\12\catcode `\$12\catcode
  `\&12\catcode `\#12\catcode `\^12\catcode `\_12\catcode `\%12\relax}%
\providecommand \@@startlink[1]{}%
\providecommand \@@endlink[0]{}%
\providecommand \url  [0]{\begingroup\@sanitize@url \@url }%
\providecommand \@url [1]{\endgroup\@href {#1}{\urlprefix }}%
\providecommand \urlprefix  [0]{URL }%
\providecommand \Eprint [0]{\href }%
\providecommand \doibase [0]{https://doi.org/}%
\providecommand \selectlanguage [0]{\@gobble}%
\providecommand \bibinfo  [0]{\@secondoftwo}%
\providecommand \bibfield  [0]{\@secondoftwo}%
\providecommand \translation [1]{[#1]}%
\providecommand \BibitemOpen [0]{}%
\providecommand \bibitemStop [0]{}%
\providecommand \bibitemNoStop [0]{.\EOS\space}%
\providecommand \EOS [0]{\spacefactor3000\relax}%
\providecommand \BibitemShut  [1]{\csname bibitem#1\endcsname}%
\let\auto@bib@innerbib\@empty
\end{thebibliography}%

\begin{widetext}

\end{widetext}

\end{document}


\title{Supplemental Material for ``Matrix Density Waves and Fractionally Charged Point Defects in Flavor Weyl Semimetals''}

\author{Shantonu Mukherjee}
\email{shantanumukherjeephy@gmail.com}
\affiliation{Department of Physics, Indian Institute of Technology Bombay,
Powai, Mumbai 400076, India}

\author{Hridis K. Pal}
\email{hridis.pal@iitb.ac.in}
\affiliation{Department of Physics, Indian Institute of Technology Bombay,
Powai, Mumbai 400076, India}

\date{\today}
\maketitle

\tableofcontents
\vspace{5mm}

We set $\hbar=1$ except where it is restored explicitly.
The Pauli matrices $\bm\rho$, $\bm\sigma$, and $\bm\tau$ act on node, Weyl
pseudospin, and flavor, respectively.

\section{Microscopic selection of the adjoint internode channel}
\label{sec:HF}

\subsection{Low-energy fields and the general coherence matrix}

Let the two opposite-chirality Weyl nodes be centered at momenta
\(\pm\mathbf b\). We write the microscopic fermion field as
\begin{equation}
c_{\alpha}(\mathbf r)
=
e^{+i\mathbf b\cdot\mathbf r}
\phi_{R\alpha}(\mathbf r)
+
e^{-i\mathbf b\cdot\mathbf r}
\phi_{L\alpha}(\mathbf r),
\label{eq:Sslowfields}
\end{equation}
where \(\alpha=1,2\) labels flavor. The fields
\(\phi_{R\alpha}\) and \(\phi_{L\alpha}\) are two-component Weyl spinors, and
their pseudospin indices will be left implicit throughout. Accordingly, a
bilinear such as
\(\phi_{R\alpha}^{\dagger}\phi_{L\beta}\) denotes contraction with the
identity matrix in pseudospin space. We restrict attention to this
pseudospin-scalar particle-hole channel.

The most general inter-node CDW order in flavor space is then described by
the \(2\times2\) matrix
\begin{equation}
M_{\alpha\beta}
=
\left\langle
\phi_{R\alpha}^{\dagger}
\phi_{L\beta}
\right\rangle
=
\frac12
\left(
B_0\tau^0+B_a\tau^a
\right)_{\alpha\beta}.
\label{eq:SMmatrix}
\end{equation}
Here \(B_0\) is the flavor-singlet component, while \(B_a\), \(a=1,2,3\),
form the flavor-adjoint component.

The physical density of flavor \(\alpha\) is
\begin{equation}
n_\alpha(\mathbf r)
=
c_\alpha^\dagger(\mathbf r)c_\alpha(\mathbf r).
\end{equation}
Using Eq.~\eqref{eq:Sslowfields}, its inter-node contribution is
\begin{equation}
\delta n_\alpha(\mathbf r)
=
e^{-2i\mathbf b\cdot\mathbf r}
M_{\alpha\alpha}
+
\text{c.c.}
\label{eq:Sdensitymod}
\end{equation}
Thus the diagonal elements of \(M\) directly describe flavor-dependent
density modulations at wave vector \(2\mathbf b\).

The total density modulation is obtained by summing over flavor,
\begin{equation}
\delta n(\mathbf r)
=
e^{-2i\mathbf b\cdot\mathbf r}
\operatorname{Tr}M
+
\text{c.c.}
\end{equation}
Since
\[
\operatorname{Tr}M=B_0,
\]
the flavor-singlet component modulates the two flavor densities in phase.
A diagonal adjoint component, for example \(B_3\), modulates the two flavor
densities oppositely. General adjoint orientations are related by global
flavor rotations and may contain off-diagonal flavor coherence. For a purely
adjoint state,
\[
\operatorname{Tr}M=0,
\]
so the leading \(2\mathbf b\) modulation of the total charge density vanishes,
while the flavor-dependent density remains spatially modulated.

\subsection{Hartree--Fock decomposition of a local repulsion}

Consider a local flavor-invariant repulsive interaction
\begin{equation}
H_U
=
\frac{U}{2}
\int d^3r\,
:n^2(\mathbf r):,
\qquad
n(\mathbf r)
=
\sum_\alpha
c_\alpha^\dagger(\mathbf r)c_\alpha(\mathbf r),
\qquad
U>0.
\label{eq:SHU}
\end{equation}
The fermion fields in Eq.~\eqref{eq:SHU} are again two-component Weyl
spinors, with their pseudospin contraction understood implicitly. The
interaction is therefore explicitly invariant under global rotations in
flavor space.

To determine how this repulsion acts on the different inter-node CDW
channels, we evaluate its expectation value in a state with a general
coherence matrix \(M\),
\begin{equation}
{
\delta F_U
=
\langle H_U\rangle_{\rm CDW}.
}
\label{eq:SHFexpectation}
\end{equation}
We retain the part quadratic in the inter-node coherence and non-oscillating after combining the two factors carrying momenta
\(\pm2\mathbf b\).

Applying Wick's theorem to the four-fermion interaction produces a direct
(Hartree) contraction and an exchange (Fock) contraction. In the
pseudospin-scalar channel defined above, the direct contribution depends on
the total density modulation and is therefore
\begin{equation}
\delta F_H
=
U
\left|
\operatorname{Tr}M
\right|^2.
\label{eq:SHartree}
\end{equation}
Explicitly,
\begin{align}
\delta F_H
=
U\Big(
|M_{11}|^2+|M_{22}|^2
+
M_{11}M_{22}^*
+
M_{22}M_{11}^*
\Big).
\end{align}

The exchange contraction carries the fermionic minus sign and gives
\begin{equation}
\delta F_F
=
-U\,\operatorname{Tr}
\left(
MM^\dagger
\right),
\label{eq:SFock}
\end{equation}
or
\begin{equation}
\delta F_F
=
-U
\left(
|M_{11}|^2
+
|M_{22}|^2
+
|M_{12}|^2
+
|M_{21}|^2
\right).
\end{equation}
Here the pseudospin contraction is already contained in the definition of
the scalar bilinear
\(\phi_{R\alpha}^{\dagger}\phi_{L\beta}\).
Any common numerical factor associated with the suppressed two-component
pseudospin trace may be absorbed into the projected coupling \(U\).

Adding the direct and exchange contributions,
\begin{equation}
\delta F_U
=
\delta F_H+\delta F_F,
\end{equation}
the terms \(|M_{11}|^2\) and \(|M_{22}|^2\) cancel, giving
\begin{align}
\delta F_U
&=
U\left(
M_{11}M_{22}^*
+
M_{22}M_{11}^*
-
|M_{12}|^2
-
|M_{21}|^2
\right)
\nonumber\\
&=
U
\left[
|\operatorname{Tr}M|^2
-
\operatorname{Tr}(MM^\dagger)
\right].
\label{eq:SHFresult1}
\end{align}

Using the decomposition
\[
M
=
\frac12
\left(
B_0\tau^0+B_a\tau^a
\right),
\]
together with
\[
\operatorname{Tr}(MM^\dagger)
=
\frac12
\left(
|B_0|^2+\sum_{a=1}^3|B_a|^2
\right),
\qquad
\operatorname{Tr}M=B_0,
\]
we obtain
\begin{equation}
{
\delta F_U
=
\frac{U}{2}
\left(
|B_0|^2
-
\sum_{a=1}^3|B_a|^2
\right).
}
\label{eq:SHFresult}
\end{equation}

Equation~\eqref{eq:SHFresult} shows directly how a repulsive microscopic
interaction acts differently on the two flavor channels. The singlet
component receives a positive contribution,
\[
\delta F_U^{(0)}
=
\frac{U}{2}|B_0|^2,
\]
whereas the adjoint component receives
\[
\delta F_U^{(\mathrm{adj})}
=
-\frac{U}{2}\sum_a|B_a|^2.
\]
Thus the local repulsion suppresses the flavor-singlet CDW while lowering
the energy of the flavor-adjoint particle-hole channel through exchange.

The quadratic free-energy contribution coming from the fermionic
susceptibility is flavor symmetric. Denoting its coefficient by \(r_0\),
\begin{equation}
\delta F_0
=
r_0\operatorname{Tr}(MM^\dagger)
=
\frac{r_0}{2}
\left(
|B_0|^2+\sum_a|B_a|^2
\right).
\label{eq:Ssusceptibility}
\end{equation}
Combining Eqs.~\eqref{eq:SHFresult} and
\eqref{eq:Ssusceptibility} gives
\begin{equation}
{
\delta F^{(2)}
=
\frac12(r_0+U)|B_0|^2
+
\frac12(r_0-U)\sum_a|B_a|^2.
}
\label{eq:Scoefficients}
\end{equation}

The repulsive interaction therefore increases the quadratic coefficient of
the singlet channel while decreasing that of the adjoint channel. As the
interaction strength is increased, the adjoint channel is the first one to
become soft. This result follows from starting with the general flavor
matrix \(M\) and comparing its singlet and adjoint components, rather than
assuming an adjoint condensate from the outset.

Equation~\eqref{eq:Scoefficients} establishes this preference for the
minimal flavor-symmetric contact model. In a microscopic Weyl material,
Bloch-state form factors, long-range Coulomb interactions, intervalley
matrix elements, lattice-scale flavor anisotropy, phonons, and other
microscopic effects may renormalize the relative channel couplings. Once
the adjoint channel has been identified as the relevant low-energy channel,
we describe it by a renormalized attractive coupling \(G>0\). Its precise
relation to the bare microscopic repulsion \(U\) depends on the microscopic
projection onto the Weyl bands.
\section{Hubbard--Stratonovich theory and the ordered manifold}
\label{sec:HS}
We start from the two-flavour Weyl action coupled to a $\mathbb{PT}$ symmetry breaking term $i \bar{\psi} \slashed{b}\gamma^5\psi$ which creates a finite node separation 
\begin{align}
    S = \int\, d^4x\, &\bigg[\bar{\psi} (\slashed{\partial} -i e \slashed{A} - i \slashed{b}\gamma^5)\psi + G \left((\bar{\psi} \mathbb{I}_4\otimes\tau^i\psi)^2 + (i \bar{\psi} \gamma^5\otimes\tau^i\psi)^2\right)\bigg]
\end{align}

To proceed we introduce two Hubbard-Stratonovich (HS) decoupling field $u^a$ and $v^a$ by introducing two gaussian integral in the generating functional 
\begin{align}
    &\int \mathcal{D} u^a  e^{-\int\, d^4x\, \left(\frac{1}{2\sqrt{G}} u^a + i\sqrt{G} S^a\right)^2}   \times \int\, \mathcal{D} v^a e^{-\int\, d^4x\, \left(\frac{1}{2\sqrt{G}} v^a + i\sqrt{G} P^a\right)^2}=\text{const}.
\end{align}
where $S^a = \bar{\psi} \mathbb{I}_4\otimes\tau^a\psi$, \, $P^a= i \bar{\psi} \gamma^5\otimes\tau^a\psi$. So the decoupled Lagrangian of the two-flavor Weyl system is given by
\begin{align}
    \mathcal{L}= \bar{\psi} (\slashed{\partial} -i e \slashed{A} - i \slashed{b}\gamma^5)\psi + i\bar{\psi} \left(\mathbb{I}_4 \otimes u^a\tau^a + i\gamma^5 \otimes v^a\tau^a\right)\psi + \frac{1}{4G} (\bm u^2 + \bm v^2).
\end{align}
The $\mathbb{PT}$-symmetry-breaking term $\bar{\psi} i\slashed{b}\gamma^5\psi$,
which encodes the separation of the two Weyl nodes in energy-momentum space, considerably complicates the direct evaluation of the fermionic effective action. In the following, we therefore focus on the commensurate charge-density-wave (CDW) configuration with ordering wave vector
$ Q_\mu=2b_\mu$. For this choice, the spatial modulation associated with the internode separation can be absorbed into the fermionic fields by an appropriate chiral transformation, thereby mapping the problem to an effectively coincident-node representation.
Using the chiral decomposition $\psi=
\begin{pmatrix}
\phi_R \\
\phi_L
\end{pmatrix},
$
the decoupled term can be written as
\begin{align}
\bar{\psi}
\left(
\mathbb I_4 u^a
+i\gamma^5 v^a
\right)
\otimes\tau^a\psi =
i\phi_L^\dagger
\left(\mathbb I_2\otimes\tau^a\right)
(u^a+i v^a)\phi_R
+
i\phi_R^\dagger
\left(\mathbb I_2\otimes\tau^a\right)
(u^a-i v^a)\phi_L .
\end{align}
It is therefore natural to introduce the complex adjoint order parameter
\begin{equation}
\Delta^a\equiv u^a+i v^a,
\qquad
\Delta^{a*}\equiv u^a-i v^a ,
\end{equation}
in terms of which we can write the decoupled term, which serves as matrix mass term of the Weyl fermions, as
\begin{align}
\bar{\psi}
\left(
\mathbb I_4 u^a
+i\gamma^5 v^a
\right)
\otimes\tau^a\psi
=
i\phi_L^\dagger
\left(\mathbb I_2\otimes\tau^a\right)
\Delta^a\phi_R
+
i\phi_R^\dagger
\left(\mathbb I_2\otimes\tau^a\right)
\Delta^{a*}\phi_L.
\end{align}
At the mean-field level, the corresponding composite order parameter is therefore proportional to the internode fermion bilinear,
\begin{equation}
\Delta^a
\sim
\phi_R^\dagger
\left(\mathbb I_2\otimes\tau^a\right)
\phi_L .
\end{equation}

Since each component of the adjoint CDW order parameter is, in general, complex, one may write
\begin{equation}
\Delta^a=\Phi^a e^{i\theta^a},
\end{equation}
where $\Phi^a$ and $\theta^a$ denote its amplitude and phase, respectively. In principle, the three adjoint components may carry independent phases. Such a general configuration considerably complicates the low-energy analysis. The effective potential, however, selects a collinear configuration in which all components share a common phase.

\section{Effective Potential for CDW}\label{appendix-A}
The effective potential after integrating out fermions is given by 
\begin{align}
     V_{\mathrm{eff}}= \frac{1}{4G}\Delta^{*a} \Delta^{a} - \int_k \ln Det[D(k)], \qquad D(k)= i\gamma^\mu k_\mu + i \left(\mathbb{I}_4 \otimes u^a\tau^a + i\gamma^5 \otimes v^a\tau^a\right).
\end{align}
The complex mass matrix can be diagonalized in the flavor space using the following transformation of the left and right Weyl spinors: $\phi_R= U_R \zeta_R,\,\, \phi_L= U_L \zeta_L$. Using these we can write the complex mass matrix as 
\begin{align}
  &\bar{\psi}  \left(\mathbb{I}_4 \otimes u^a\tau^a + i\gamma^5 \otimes v^a\tau^a\right) \psi \nonumber \\ &= 
    i\phi_L^\dagger
\left(\mathbb I_2\otimes\tau^a\right)
\Delta^a\phi_R
+
i\phi_R^\dagger
\left(\mathbb I_2\otimes\tau^a\right)
\Delta^{a*}\phi_L \\ & 
=
    i \zeta^\dagger_L U^\dagger_L \mathbb{I}_2 \otimes M U_R \zeta_R + i \zeta^\dagger_R U^\dagger_R \mathbb{I}_2 \otimes M^\dagger U_L \zeta_L
\end{align}
where we have written $M= \tau^a \Delta^a,\,\,\, M^\dagger= \tau^a \Delta^{*a}$. As $U_L,\, U_R$ diagonalizes $M$ we can write
\begin{align}
   U^\dagger_R \mathbb{I}_2 \otimes M U_L = \mathbb{I}_2\otimes\Sigma,\quad U^\dagger_L \mathbb{I}_2 \otimes \Delta^{a*}\tau^a U_R = \mathbb{I}_2\otimes\Sigma,
    \Sigma = \begin{pmatrix}
        m_+ && 0\\
        0 && m_{-}
    \end{pmatrix}
\end{align}
Therefore, the diagonalized mass term is 
\begin{equation}
     i \bar{\zeta} \mathbb{I}_4 \otimes \Sigma \zeta
\end{equation}
This will give us the following form of the Dirac operator $D = i \slashed{k}\otimes \mathbb{I}_2 + i \mathbb{I}_4 \otimes \Sigma$. So the determinant of this Dirac operator is given by
\begin{align}
    & det_{D,f}(D)= det_D (i \slashed{k} + i m_+ \mathbb{I}_4 )\, det_D (i \slashed{k} + i m_- \mathbb{I}_4 ) = (k^2 + m^2_+)^2 (k^2 + m^2_-)^2
\end{align}
As $M = U^\dagger_L \Sigma U_R$, $M^\dagger M = U^\dagger_R \Sigma^2 U_R$. So $\Sigma^2$ is the diagonalized form of $M^\dagger M$ and $m^2_+,\, m^2_-$ are it's eigen values. Again, 
\begin{equation}
    M^\dagger M = \Delta^{a*} \Delta^a + i \vec{\tau}\cdot (\vec{\Delta}^* \times \vec{\Delta}).
\end{equation}
Designating $\varrho =  \Delta^{a*} \Delta^a $, $c= |i (\vec{\Delta}^* \times \vec{\Delta})|$, it is easy to show that the eigen value of $\Sigma^2$ are $m^2_{\pm}= \varrho \pm c$. This will finally give us
\begin{equation}
    det_{D,f}(D) = ((k^2+\varrho)^2 - c^2)^2.
\end{equation}
Using this expression of the determinant we determine the effective potential
\begin{equation}
    V_{\mathrm{eff}}= \frac{\varrho}{4G} - 2 \int \frac{d^4k}{(2\pi)^4} \ln[(k^2 + \varrho)^2 - c^2].
\end{equation}
For a spatially uniform CDW configuration, this one-loop effective potential satisfies
\begin{equation}
V_{\mathrm{eff}}(\varrho,c)- V_{\mathrm{eff}}(\varrho,0)=
-2
\int\frac{d^4k}{(2\pi)^4}
\ln\left[
1-\frac{c^2}{(k^2+\varrho)^2}
\right].
\label{eq:Veff-collinear}
\end{equation}
Since $
0\leq
\frac{c^2}{(k^2+\rho)^2}
\leq1
$
and $-\ln(1-x)\geq0$ for $0\leq x<1$, Eq.~\eqref{eq:Veff-collinear} implies
\begin{equation}
V_{\mathrm{eff}}(\varrho,c)
\geq
V_{\mathrm{eff}}(\varrho,0).
\end{equation}
The minimum of the effective potential therefore occurs at
\begin{equation}
c=0,
\end{equation}
or equivalently,
\begin{equation}
\vec{\Delta}^{*}\times\vec{\Delta}=0.
\end{equation}
Writing $\vec{\Delta}= \bm u +i \bm v$, this condition implies that $\vec{u}$ and $\vec{v}$ are parallel. Consequently, the adjoint order parameter can be expressed in the collinear form
\begin{equation}
\Delta^a=\Phi^a e^{i\theta}= \Phi\, n^a e^{i\theta},
\label{eq:collinear cdw}
\end{equation}
where $\Phi^a$ is a real-valued adjoint amplitude field and $\theta$ is a common $U(1)$ phase. In the remainder of this work, we restrict our analysis to this energetically favored collinear CDW phase.\\

At fixed nonzero $\Phi$, Eq.~\eqref{eq:collinear cdw} is unchanged by
\begin{equation}
(\mathbf n,\theta)\longrightarrow(-\mathbf n,\theta+\pi).
\label{eq:SZ2}
\end{equation}
The vacuum manifold is consequently
\begin{equation}
\mathcal M=\frac{S^2\times S^1}{\mathbb Z_2},
\qquad \pi_2(\mathcal M)=\mathbb Z.
\label{eq:Smanifold}
\end{equation}
 \subsection{$\Phi^a$ is an adjoint field}
The field $\Phi^a$ transforms in the adjoint representation of the internal $SU(2)$ flavor symmetry. Under the global flavor transformation
\begin{equation}
\phi_{L,R}
\rightarrow
e^{i\alpha^a\tau^a}
\phi_{L,R},
\end{equation}
the matrix-valued order parameter $
\Phi\equiv\Phi^a\tau^a
$
transforms as
\begin{equation}
\Phi\rightarrow U\Phi U^\dagger,
\qquad
U=e^{\,i\alpha^a\tau^a}.
\end{equation}
For an infinitesimal transformation, this gives
\begin{equation}
\delta\Phi^a
=
-2\epsilon^{abc}\alpha^b\Phi^c,
\end{equation}
up to the normalization convention chosen for the $SU(2)$ generators. Thus $\vec{\Phi}$ transforms as a three-component vector under the adjoint $SO(3)$ action associated with the internal $SU(2)$ flavor symmetry. The corresponding rotation takes place in flavor space rather than in physical coordinate space.
\subsection{Chiral transformation of Weyl spinors and transformed action}
Because the CDW order parameter couples fermionic states originating from Weyl nodes separated by $2b_\mu$, it carries the characteristic internode modulation
\begin{equation}
\Delta^a(x)
=
\Phi^a(x)
e^{\,i[\theta(x)-Q\cdot x]},
\qquad
Q_\mu=2b_\mu.
\end{equation}
The slowly varying phase $\theta(x)$ describes the collective CDW phase mode, whereas the factor $e^{iQ\cdot x}$ accounts for the microscopic internode modulation.
We now perform the chiral rotation
\begin{equation}
\psi
\rightarrow
\exp\left[
-\frac{i}{2}
\gamma^5
\left(\theta-Q\cdot x\right)
\right]\psi .
\end{equation}
For $Q_\mu=2b_\mu$, this transformation removes the explicit node-separation term from the fermionic kinetic operator and simultaneously removes the phase of the CDW mass term. The spacetime dependence of the transformation generates an axial-vector coupling proportional to $\partial_\mu\theta$, while the non-invariance of the fermionic path-integral measure produces the anomalous axion contribution. The resulting action can therefore be written as
\begin{widetext}
\begin{equation}
\label{decoupled}
\begin{split}
S
=
\int d^4x\,
\Bigg[
N_f
\frac{\theta-Q\cdot x}{2}
\frac{e^2}{16\pi^2}
\,i\varepsilon^{\mu\nu\rho\lambda}
F_{\mu\nu}F_{\rho\lambda}
+
\frac{1}{4G}
\Phi^{a}\Phi^a
\
&
+
\bar{\psi}
\left(
\slashed{\partial}
-ie\slashed A
-\frac{i}{2}
\gamma^\mu\gamma^5\partial_\mu\theta
\right)
\psi
+
i\bar{\psi}
\left(
\mathbb{I}_4 \otimes \Phi^a\tau^a
\right)
\psi
\Bigg].
\end{split}
\end{equation}
\end{widetext}
Here the rapidly varying internode phase has been transferred into the anomalous axion term, while the CDW order parameter entering the fermionic mass sector is reduced to the real adjoint amplitude $\Phi^a$. A nonzero expectation value
\begin{equation}
\langle\Phi^a\rangle\neq0
\end{equation}
therefore signals the formation of the adjoint CDW phase. In the following section, we determine the condition under which such a nonzero amplitude becomes energetically favorable.

\subsection{Gap equation and critical coupling}
To determine the critical condition for the spontaneous symmetry breaking into the CDW phase we shall obtain the gap equation for the CDW order parameter $\Phi^a$. To do this we shall ignore the electromagnetic fields. With these assumptions in mind, let us start the derivation with the following Lagrangian
\begin{equation}
    \mathcal{L}= \frac{1}{4G} (\Phi^{a} \Phi^a ) + \bar{\psi} \left(\gamma^\mu \partial_\mu + i \mathbb{I}_4\otimes \Phi^a \tau^a\right)\psi.
\end{equation}
In momentum space the action takes the following form
\begin{align}
   & S  = \sum_{k,k^\prime} \bar{\psi}_{k^\prime} M_{k,\, k^\prime}\psi_k + \frac{1}{4G L^4} \sum_k \Phi^a_k \Phi^a_{-k},  \\ &
   M_{k,\, k^\prime} = \frac{1}{(L^4)^2}\left(\gamma^\mu\otimes \mathbb{I}_2\, (-i k_\mu)\, \delta_{k,\, k^\prime} L^4 + i\mathbb{I}_4\otimes \Phi^a_{k,\,k^\prime} \tau^a\right),
\end{align}
where $L^4$ is the volume of the  discretised spacetime. We shall obtain the effective action for the order parameter $\Phi^a$ by integrating out the fermions using the generating functional $\mathcal{Z}$ as follows
\begin{equation}
    \mathcal{Z}= \int \mathcal{D}\bar{\psi} \mathcal{D}\psi \, \exp{ -\sum_{k,k^\prime} \bar{\psi}_{k^\prime} M_{k,\, k^\prime}\psi_k } \sim e^{\Tr{\ln{M}}}.
\end{equation}
This will lead to the following effective action for $\Phi^a$
\begin{equation}
    S_{\Phi^a}= \frac{1}{4G L^4} \sum_k \Phi^a_k \Phi^a_{-k} - \Tr{\ln{M}}.
\end{equation}
For a uniform collinear saddle, $\varrho=\Phi_0^2$ and $c=0$. Let us now derive the equation of motion of the order parameter field $\Phi^a$
\begin{equation}
    \frac{\delta\, S_{\Phi^a} }{\delta \Phi^a_q} = 0.
\end{equation}
This will lead to 
\begin{equation}
    \frac{1}{L^4 2G} \Phi^a{-q} - \sum_{p} \tr{\left(\frac{1}{M} \times \frac{\delta }{\delta \Phi^a_q} M\right)}_{p,p}=0,
\end{equation}
where the $\Tr$ represents the trace of $M$ which is a matrix in the momentum as well as in spin-flavor space, whereas $\mathbf{\tr}$ represents only trace over spin-flavor space. Performing the derivative $\frac{\delta }{\delta \Phi^a_q} M$ we get finally
\begin{equation}\label{gap equation}
    \frac{1}{L^4 2G} \Phi^a_{-q} - \sum_{p} \tr{\left(\frac{1}{M}\right)_{p,\, p+q} \frac{i\mathbb{I}_4\otimes \tau^a}{(L^4)^2}   }=0
\end{equation}
This is the required equation for the order-parameter field $\Phi^a$. We now remember that in the collinear phase the form of the CDW order is given by $\Delta^a\sim \Phi^a\, e^{i\theta - i Q\cdot r}$ and therefore the CDW order have a wave like modulation of the form~$e^{- i Q\cdot r}$. Therefore for the minimal condition, the amplitude part can be considered as a $Q$ dependent constant i.e. $q=0$. This will make the Matrix $M$ diagonal and so as $1/M$. This will lead to the following simplified form of the equation Eq~\eqref{gap equation}

    \begin{equation}
    \frac{1}{2G}\Phi^a_0 - \int\,\frac{d^4k}{(2\pi)^4} \tr{\frac{(i\mathbb{I}_4\otimes \tau^a)}{\gamma^\mu\otimes \mathbb{I}_2\, (-i k_\mu) + i\mathbb{I}_4\otimes \Phi^b_0\tau^b} } = 0.
\end{equation}

Here we have used $\Phi^a_{q=0}= L^4 \Phi_0^a$. Performing the integral explicitly with the 4-momentum cutoff scale $\Lambda$ which is much larger than the gap $\sqrt{\Phi^a_0\Phi^a_0}$ we get
\begin{equation}
    \frac{N_f \Lambda^2}{4\pi^2} - \frac{1}{2G} = \frac{N_f (\Phi^a_0)^2}{4\pi^2} \ln{\left(\frac{\Lambda^2}{(\Phi^a_0)^2} +1\right)}.
\end{equation}
\subsection{Leading derivative expansion}
We now derive the leading terms in the low-energy effective action by integrating out the fermionic degrees of freedom. The four-momentum integrals are regularized using an ultraviolet cutoff $\Lambda$, which represents the range of validity of the low-energy continuum description. In the physical theory considered below, no additional flavor-singlet fermion mass is introduced.

Denoting collectively the couplings to the adjoint amplitude, electromagnetic field, and axial field by
\begin{equation}
\mathcal{V}
=
e\slashed{A} + \slashed{A}^{5}\gamma^5 - \Phi^a \tau^a,
\end{equation}
the fermionic contribution to the effective action can be expanded as
\begin{align}
  &  -\Tr\ln\left(G_0^{-1}- i\mathcal V\right)\nonumber
\\ &=
-\Tr\ln G_0^{-1}
+\Tr(G_0i \mathcal V)
+\frac{1}{2}\Tr(G_0 i\mathcal V G_0 i\mathcal V)
+\cdots ,
\end{align}
where
\begin{equation}
G_0(k)=\frac{-i}{\slashed{k} -m}
\end{equation}
is the massless fermion propagator. The factor $1/2$ appearing in the two-leg contribution follows directly from the expansion of the logarithm. In the adjoint sector we use
\begin{equation}
\Tr_f(T^aT^b)=N_f\,\delta^{ab},
\end{equation}
where $N_f=2$ for the two-flavor model considered here.

Introducing a Feynman parameter $x$, the quadratic contribution can be decomposed into the amplitude, electromagnetic, axial, and mixed vector--axial sectors. Schematically,
\begin{align}
S_{\mathrm{eff}}^{(2)}
={}&
S_{\Phi}^{(2)}
+
S_{AA}^{(2)}
+
S_{A_5A_5}^{(2)}
+
S_{AA_5}^{(2)} .
\end{align}
For the symmetry-preserving background considered here, the mixed vector--axial two-point function vanishes, and we discuss the remaining terms separately below.

\subsection{Effective action for the adjoint amplitude $\Phi^a$}\label{sub:eff action}

The quadratic contribution involving two insertions of $\Phi^a$ is
\begin{align}
S_\Phi^{(2)}=
\frac{4N_f}{2}
\int\frac{d^4p}{(2\pi)^4},
\Phi^a(p)
\int_0^1dx
\int\frac{d^4k}{(2\pi)^4}
\frac{(k\cdot(k-p)) - m^2}
{\left[(k-px)^2+Q^2\right]^2}
\Phi^a(-p),
\end{align}
where the factor $4$ originates from the Dirac trace, while $N_f$ follows from the flavor trace. For massless fermions,
\begin{equation}
Q^2=x(1-x)p^2.
\end{equation}
After shifting the loop momentum according to
\begin{equation}
k^\prime=k-px,
\end{equation}
the numerator becomes
\begin{equation}
k\cdot(k-p)
\longrightarrow
(k^\prime)^2-Q^2,
\end{equation}
after terms odd in $k^\prime$ are discarded. The relevant loop integral is therefore
\begin{equation}
I(Q^2)
=
\int^\Lambda\frac{d^4k^\prime}{(2\pi)^4}
\frac{(k^\prime)^2-Q^2}
{\left[(k^\prime)^2+Q^2\right]^2}.
\end{equation}
Evaluating the integral with a four-dimensional momentum cutoff gives

    \begin{equation}
I(Q^2)
=
\frac{1}{16\pi^2}
\left[
\Lambda^2
+2Q^2
-3Q^2
\ln\left(1+\frac{\Lambda^2}{Q^2}\right)
-\frac{2Q^4}{Q^2+\Lambda^2}
\right].
\end{equation}

Combining this result with the Hubbard--Stratonovich term, the quadratic effective action becomes
\begin{widetext}
\begin{align}
S_\Phi^{(2)}
=
\int\frac{d^4p}{(2\pi)^4}\,
\Phi^a(p)
\int_0^1dx,
\Bigg[
\frac{1}{4G}
-\frac{N_f\Lambda^2}{8\pi^2}
-\frac{N_fQ^2}{4\pi^2}
+\frac{3N_fQ^2}{8\pi^2}
\ln\left(1+\frac{\Lambda^2}{Q^2}\right)
+\frac{N_fQ^4}
{4\pi^2(Q^2+\Lambda^2)}
\Bigg]
\Phi^a(-p).
\label{eq:Phi-quadratic-action}
\end{align}
\end{widetext}

For $p^2\ll\Lambda^2$, the last term in Eq.~\eqref{eq:Phi-quadratic-action} contributes only at order $p^4/\Lambda^2$ and may be neglected in the leading derivative expansion. Using $Q^2=x(1-x)p^2$, together with
\begin{equation}
\int_0^1dx,x(1-x)=\frac{1}{6},
\qquad
\int_0^1dx\,x(1-x)\ln[x(1-x)]
=
-\frac{5}{18},
\end{equation}
we obtain
\begin{widetext}
    \begin{align}
S_\Phi^{(2)}
=\int\frac{d^4p}{(2\pi)^4},
\Phi^a(p)
\left[
\frac{1}{4G}
-\frac{N_f\Lambda^2}{8\pi^2}
+
\frac{N_fp^2}{16\pi^2}
\left(
\ln\frac{\Lambda^2}{p^2}+1
\right)
+\mathcal O\left(\frac{p^4}{\Lambda^2}\right)
\right]
\Phi^a(-p).
\label{eq:Phi-small-p}
\end{align}
\end{widetext}

It is useful to rewrite Eq.~\eqref{eq:Phi-small-p} in the conventional quadratic form
\begin{equation}
S_\Phi^{(2)}
=
\frac{1}{2}
\int\frac{d^4p}{(2\pi)^4},
\Phi^a(p)
\left[
m_\Phi^2+\rho_\Phi(p)p^2
\right]
\Phi^a(-p),
\end{equation}
where
\begin{equation}
{
m_\Phi^2
= \frac{1}{2G} -
\frac{N_f\Lambda^2}{4\pi^2}
}
\end{equation}
and
\begin{equation}
{
\rho_\Phi(p)
=
\frac{N_f}{8\pi^2}
\left[
\ln\left(\frac{\Lambda^2}{p^2}\right)+1
\right].
}
\end{equation}
The sign change of $m_\Phi^2$ determines the instability toward the formation of the adjoint CDW and gives the critical condition
\begin{equation}
\frac{1}{2G_c}
=
\frac{N_f\Lambda^2}{4\pi^2},
\end{equation}
in agreement with the corresponding mean-field gap equation.

The momentum-dependent contribution in Eq.~\eqref{eq:Phi-small-p} represents the leading stiffness of the adjoint CDW field. Since the underlying fermions are gapless in the symmetric phase, the stiffness exhibits a logarithmic dependence on the external momentum. We therefore define the stiffness at a finite characteristic momentum scale $\mu$,
\begin{equation}
{
\rho_\Phi(\mu)
=
\frac{N_f}{8\pi^2}
\left[
\ln\left(\frac{\Lambda^2}{\mu^2}\right)+1
\right].
}
\label{eq:rho-Phi}
\end{equation}
Here $\mu$ specifies the momentum scale at which the long-wavelength response is probed and does not correspond to an additional fermionic mass. The leading local contribution to the effective action may consequently be written as
\begin{equation}
S_{\Phi\,\mathrm{grad}}
=
\frac{\rho_\Phi(\mu)}{2}
\int d^4x\,
(\partial_\mu\Phi^a)(\partial_\mu\Phi^a).
\end{equation}
Importantly, $\Phi^a$ in the above expression denotes the full adjoint order-parameter field rather than a small fluctuation around a fixed uniform condensate. The resulting effective theory therefore allows spatial configurations in which the magnitude of $\Phi^a$ varies substantially, including configurations for which $\Phi^a$ vanishes locally, as required for the monopole textures considered in this work.

\subsection{Effective action for $A_\mu$ and $\theta$}

The two-leg fermionic diagram also generates the electromagnetic response. After Feynman parametrization and the shift of loop momentum, the vector polarization contains the structure
\begin{widetext}
    \begin{align}
S_{AA}^{(2)}
=
\frac{4\,e^2 N_f}{2} \int_0^1dx
\int\frac{d^4k}{(2\pi)^4}
\frac{d^4p}{(2\pi)^4}\,
A_\mu(p)
\Bigg[
\frac{2x(1-x)
\left(
p^2\delta_{\mu\nu}-p_\mu p_\nu
\right)}
{(k^2+Q^2)^2}
+
\frac{2k_\mu k_\nu}{(k^2+Q^2)^2}
-
\frac{\delta_{\mu\nu}}{k^2+Q^2}
\Bigg]
A_\nu(-p),
\label{eq:vector-polarization}
\end{align}
\end{widetext}
where the overall coupling-dependent prefactors are understood.

A sharp momentum cutoff applied directly to the individual terms in Eq.~\eqref{eq:vector-polarization} can produce an apparent photon mass because such a cutoff does not, in general, preserve momentum-shift invariance. To implement the ultraviolet cutoff while preserving the Ward identity, we use the Schwinger proper-time representation
\begin{equation}
\frac{1}{k^2+Q^2}
=
\int_{1/\Lambda^2}^{\infty}ds\,
e^{-s(k^2+Q^2)},
\end{equation}
and
\begin{equation}
\frac{1}{(k^2+Q^2)^2}
=
\int_{1/\Lambda^2}^{\infty}ds\,
s\,e^{-s(k^2+Q^2)}.
\end{equation}
The Gaussian integral satisfies
\begin{equation}
\int_k
k_\mu k_\nu e^{-sk^2}
=
\frac{\delta_{\mu\nu}}{2s}
\int_k e^{-sk^2}.
\end{equation}
Consequently,
\begin{widetext}
    \begin{align}
\int_k
\left[
\frac{2k_\mu k_\nu}{(k^2+Q^2)^2}
-
\frac{\delta_{\mu\nu}}{k^2+Q^2}
\right]
=
\int_{1/\Lambda^2}^{\infty}ds
\int_k
\left[
2s k_\mu k_\nu-\delta_{\mu\nu}
\right]
e^{-s(k^2+Q^2)}=0.
\end{align}
\end{widetext}
Thus the potentially gauge-noninvariant momentum-independent contribution cancels identically, and the electromagnetic polarization remains transverse,
\begin{equation}
\Pi_{AA}^{\mu\nu}(p)
=
\left(
p^2\delta^{\mu\nu}-p^\mu p^\nu
\right)
\Pi_A(p^2).
\end{equation}
The leading local electromagnetic contribution may therefore be expressed as
\begin{equation}
\mathcal L_A
=
-\frac{Z_A}{4}
F_{\mu\nu}F^{\mu\nu}.
\end{equation}

The phase field $\theta$ enters through the axial-vector combination
\begin{equation}
A_\mu^5=\frac{1}{2}\partial_\mu\theta.
\end{equation}
Unlike the electromagnetic field, $A_\mu^5$ is not protected by an electromagnetic Ward identity. In the ordered CDW phase, the axial response evaluated in the gapped background generates the leading phase-stiffness contribution
\begin{equation}
\mathcal L_\theta
=
\frac{\rho_\theta}{2}
(\partial_\mu\theta)^2.
\end{equation}
The coefficient $\rho_\theta$ is determined by the corresponding axial two-point function in the ordered state. 
Collecting the leading two-derivative contributions, together with the anomalous axion term generated by the chiral transformation, we obtain
\begin{widetext}
    \begin{align}
S_{\mathrm{eff}}
=
\int d^4x\,
\Bigg[
\frac{\rho_\Phi(\mu)}{2}
(\partial_\mu\Phi^a)^2
+
\frac{m_\Phi^2}{2}
\Phi^a\Phi^a
+
\frac{\rho_\theta}{2}
(\partial_\mu\theta)^2
-
\frac{Z_A}{4}
F_{\mu\nu}F^{\mu\nu}
+
N_f
\frac{\theta-Q\cdot x}{2}
\frac{e^2}{16\pi^2}
\,i\varepsilon^{\mu\nu\rho\lambda}
F_{\mu\nu}F_{\rho\lambda}
+\cdots
\Bigg].
\label{eq:quadratic-effective-action}
\end{align}
\end{widetext}
Higher powers of the adjoint field and higher-derivative operators arise from higher-leg diagrams and from subleading terms in the momentum expansion. In particular, the leading stabilizing interaction may be written as
\begin{equation}
\frac{\lambda}{4}
\left(\Phi^a\Phi^a\right)^2,
\end{equation}
which, together with Eq.~\eqref{eq:quadratic-effective-action}, defines the long-wavelength Landau--Ginzburg theory used in this work to study spatially nonuniform adjoint-CDW configurations.

\section{Hedgehog texture and its topology}
\label{sec:hedgehog}

Hold the common phase $\theta$ fixed and write
\begin{equation}
\bm\Phi(\mathbf r)=\Phi_0 f(r)\mathbf n(\mathbf r),
\qquad f(0)=0,\quad f(\infty)=1.
\label{eq:Stexture}
\end{equation}
On a large sphere, $\mathbf n$ defines a map
$S^2_{\rm space}\to S^2_{\rm flavor}$ with degree
\begin{equation}
N=\frac{1}{4\pi}\int d\vartheta d\varphi\,
\mathbf n\cdot(\partial_\vartheta\mathbf n\times
\partial_\varphi\mathbf n).
\label{eq:Swinding}
\end{equation}

The effective field theoretic action (Eq~\eqref{eq:quadratic-effective-action} upto quartic order terms) generates a vacuum expectation value for $\Phi_0 (= \sqrt{-\tfrac{m^2_\Phi}{\lambda}})$ which makes the appearance of topological defect possible.  Inserting this functional form $\bm\Phi(\mathbf r)=\Phi_0 f(r)\mathbf n(\mathbf r)$ in the EFT action (Eq~\eqref{eq:quadratic-effective-action}) one can show that the radial function $f(r)$ obeys the following equation of motion
\begin{equation}
    \grad^2 f(r) - \frac{2}{r^2} f(r) + \frac{m^2_\Phi }{\rho_\Phi} f(r) \left(  f(r)^2 -1\right)=0.
\end{equation}
This equation has smooth solution for the above mentioned boundary condition. Therefore our EFT description admits global monopole solution. The most appropriate boundary behavior is produced by the function $f(r)= \tanh(r/ R)$, where $R=-\tfrac{2\rho_\Phi}{(m^2_\Phi)}$ is the characterize size of the monopole.\\

For the elementary hedgehog the above mentioned map between the vacuum manifold ($S^2_{\rm flavor}$) and spatial infinity ($S^2_{\rm space}$) is expressed by the relation $\mathbf n=\hat{\mathbf r}$ and has $N=1$.
For $\mathbf n=\hat{\mathbf r}$,
$(\partial_i n^a)(\partial_i n^a)=2/r^2$. The asymptotic gradient energy is
\begin{equation}
E_{\rm grad}(R)=\frac{\rho_n}{2}
\int_\xi^R d^3r\,(\partial_i n^a)^2
=4\pi\rho_n(R-\xi).
\label{eq:Sglobalenergy}
\end{equation}
The energy of the global monopole therefore linearly rises with the monopole core size $R$. A finite-energy configuration requires
an antimonopole, a compensating boundary texture, or another infrared cutoff.
This energetic statement does not remove the local core spectrum when the
compensating object is far away.

The asymptotic mass matrix $\hat M=\mathbf n\cdot\bm\tau$ has projectors
\begin{equation}
P_\pm=\frac12(1\pm\mathbf n\cdot\bm\tau).
\label{eq:Sprojectors}
\end{equation}
The two eigenline bundles have curvatures
\begin{equation}
(\mathcal F_\pm)_{\vartheta\varphi}
=i\operatorname{Tr}
P_\pm[\partial_\vartheta P_\pm,\partial_\varphi P_\pm]
=\mp\frac12\mathbf n\cdot
(\partial_\vartheta\mathbf n\times\partial_\varphi\mathbf n),
\label{eq:Sberrycurv}
\end{equation}
and Chern numbers $C_\pm=\mp N$. These characterize the matrix mass. They are
not Chern numbers of two energetically separated quasiparticle bands: the
uniform Dirac energies depend on $\Phi_0^2$ and are degenerate with respect to
the sign of the mass eigenvalue.
\section{One dimensional chiral modes bound to an Alice string}
\label{sec:alice-index}

We now determine the fermionic modes bound to an Alice string in the
two-flavor adjoint charge-density-wave state. Consider a straight string
oriented along the $z$ direction. Far from the string core, the internode
mass matrix may be written as
\begin{equation}
\Delta(\varphi)
=
\Phi_0 e^{i\theta(\varphi)}
\,\mathbf n(\varphi)\cdot\boldsymbol{\tau},
\label{eq:alice-mass}
\end{equation}
where $\varphi$ denotes the polar angle in the transverse $(x,y)$ plane,
$\theta$ is the density-wave phase, and $\mathbf n$ specifies the
orientation of the real adjoint mass in the vacuum manifold
$SO(3)/SO(2)\simeq S^2$.

For a straight line defect, the fermionic problem separates into a
two-dimensional Dirac problem in the plane transverse to the string and
free propagation along the $z$ direction. A normalizable zero mode of the
transverse Dirac operator gives rise to a one-dimensional mode localized
near the string, with low-energy dispersion
\begin{equation}
E(k_z)=\pm v_z k_z.
\end{equation}
The net chirality of these modes is determined by the index of the
transverse Dirac operator,
\begin{equation}
\nu_{\rm line}
=
N_{\rm R}-N_{\rm L}
=
\frac{1}{2\pi i}
\oint d\varphi\,
\partial_\varphi
\log \det\Delta(\varphi),
\label{eq:alice-line-index}
\end{equation}
or equivalently
\begin{equation}
\nu_{\rm line}
=
\operatorname{wind}\!\left[\det\Delta\right].
\end{equation}
Here $N_{\rm R}$ and $N_{\rm L}$ denote the numbers of right- and
left-moving modes along the defect, respectively
\cite{jackiwrossi,weinberg1981}. For the adjoint mass in Eq.~\eqref{eq:alice-mass}, the matrix
$\mathbf n\cdot\boldsymbol{\tau}$ has eigenvalues $+1$ and $-1$, since
$\mathbf n^2=1$. The two eigenvalues of the full mass matrix are therefore
\begin{equation}
\lambda_+
=
+\Phi_0 e^{i\theta},
\qquad
\lambda_-
=
-\Phi_0 e^{i\theta}.
\end{equation}
Consequently,
\begin{equation}
\det\Delta
=
\lambda_+\lambda_-
=
-\Phi_0^2 e^{2i\theta}.
\label{eq:det-adjoint-mass}
\end{equation}
The phase of the determinant is therefore
\begin{equation}
\arg\det\Delta
=
2\theta+\pi,
\end{equation}
where the constant shift by $\pi$ does not affect its winding number. The elementary Alice string is possible because the adjoint order
parameter obeys the identification
\begin{equation}
(\theta,\mathbf n)
\sim
(\theta+\pi,-\mathbf n).
\label{eq:alice-identification}
\end{equation}
Indeed,
\begin{equation}
e^{i(\theta+\pi)}
\left(-\mathbf n\cdot\boldsymbol{\tau}\right)
=
e^{i\theta}
\mathbf n\cdot\boldsymbol{\tau}.
\end{equation}
Hence the physical mass matrix is single valued even though the
density-wave phase and the adjoint orientation individually change after
one circuit around the defect. For the elementary Alice string,
\begin{equation}
\theta(2\pi)-\theta(0)=\pi,
\qquad
\mathbf n(2\pi)=-\mathbf n(0).
\label{eq:alice-boundary}
\end{equation}
Thus the density-wave phase carries only half of the usual $2\pi$ vortex
winding.

Although $\theta$ winds only by $\pi$, Eq.~\eqref{eq:det-adjoint-mass}
shows that the phase of the determinant winds by
\begin{equation}
\delta (\arg\det\Delta)
=
2\delta\theta
=
2\pi.
\end{equation}
The corresponding index is therefore
\begin{equation}
{
\nu_{\rm line}
=
\operatorname{wind}\!\left[\det\Delta\right]
=
1.
}
\label{eq:alice-index-one}
\end{equation}
The Alice string consequently carries a single unpaired chiral electronic
channel. The index fixes only the difference $N_{\rm R}-N_{\rm L}$;
additional pairs of counterpropagating modes are not excluded by the
index theorem, but the net chirality is topologically fixed to unity.

It is useful to compare the Alice string with an ordinary $2\pi$
density-wave vortex in the same two-flavor system. If the flavor
orientation is held fixed while
\begin{equation}
\delta\theta=2\pi,
\end{equation}
then
\begin{equation}
\delta(\arg\det\Delta)
=
4\pi,
\end{equation}
and hence
\begin{equation}
\nu_{\rm line}=2.
\end{equation}
A conventional full vortex of the two-flavor adjoint CDW therefore carries
two net chiral modes, whereas the elementary Alice string carries one.

For comparison, in a system containing a single Weyl pair, the internode
CDW mass is a complex scalar,
\begin{equation}
\Delta_{\rm scalar}
=
m e^{i\theta}.
\end{equation}
A conventional axion string requires a full $2\pi$ winding of $\theta$,
for which
\begin{equation}
\operatorname{wind}\Delta_{\rm scalar}=1,
\end{equation}
and therefore supports a single chiral channel. The two-flavor Alice
string has the same electronic index even though its density-wave phase
winds only by $\pi$:
\begin{equation}
\begin{aligned}
\text{single Weyl pair:}
&\qquad
\delta\theta=2\pi
\quad\Longrightarrow\quad
\nu_{\rm line}=1,
\\[2mm]
\text{two-flavor Alice string:}
&\qquad
\delta\theta=\pi,
\quad
\mathbf n\rightarrow-\mathbf n
\quad\Longrightarrow\quad
\nu_{\rm line}=1.
\end{aligned}
\end{equation}

The role of the flavor sector is therefore not to generate the fermionic
index directly. Instead, the nontrivial identification
\eqref{eq:alice-identification} allows a $\pi$ density-wave vortex to form
a closed physical configuration. The electronic index is then determined
by the winding of the full matrix-valued mass through
$\det\Delta$. In this sense, the flavor degree of freedom converts the
conventional axion string into a half-quantum defect without reducing its
topologically protected chiral index.
\section{Exact zero mode of the elementary hedgehog}
\label{sec:zeromode}
Consider the continuum Hamiltonian
\begin{equation}
H_{\rm h}
=
-i v_F\,
\rho_z\otimes
\bigl(\bm{\sigma}\cdot\bm{\nabla}\bigr)
\otimes\mathbb I_\tau
+
\Phi(r)\,
\rho_x\otimes\mathbb I_\sigma\otimes
\bigl(\hat{\bm r}\cdot\bm{\tau}\bigr),
\label{eq:Hhedgehog}
\end{equation}
with
\begin{equation}
\Phi(r)=\Phi_0 f(r),
\qquad
f(0)=0,
\qquad
f(r\to\infty)=1,
\qquad
\Phi_0>0.
\label{eq:profile}
\end{equation}
Here $\bm{\sigma}$ acts on the Weyl pseudospin, $\bm{\tau}$ on flavor,
and $\bm{\rho}$ on the pair of opposite-chirality Weyl nodes.
The Hamiltonian anticommutes with
\begin{equation}
\Gamma
=
\rho_y\otimes\mathbb I_\sigma\otimes\mathbb I_\tau,
\end{equation}
since
\begin{equation}
\{\rho_y,\rho_z\}=0,
\qquad
\{\rho_y,\rho_x\}=0.
\end{equation}
Hence
\begin{equation}
{
\{H_{\rm h},\Gamma\}=0.
}
\label{eq:grading}
\end{equation}

At zero energy, the kernel of $H_{\rm h}$ is invariant under $\Gamma$.
Therefore a zero mode may be chosen with definite grading,
\begin{equation}
\Gamma\Psi_\eta=\eta\Psi_\eta,
\qquad
\eta=\pm1.
\label{eq:eta}
\end{equation}

Multiplying $H_{\rm h}\Psi_\eta=0$ by
$\rho_z\otimes\mathbb I_\sigma\otimes\mathbb I_\tau$ and using
$\rho_z\rho_x=i\rho_y$ gives
\begin{equation}
\left[
v_F
\bigl(\bm{\sigma}\cdot\bm{\nabla}\bigr)\otimes\mathbb I_\tau
-
\eta\,\Phi(r)\,
\mathbb I_\sigma\otimes
\bigl(\hat{\bm r}\cdot\bm{\tau}\bigr)
\right]
\psi_\eta(\bm r)
=0,
\label{eq:reduced-zero}
\end{equation}
where the explicit $\rho$-space factor has been suppressed.

\subsection{Grand angular momentum}

The hedgehog is invariant under simultaneous rotations of real space,
Weyl pseudospin, and flavor. The conserved grand angular momentum is
\begin{equation}
{
\bm K
=
\bm L
+
\frac{\bm\sigma}{2}
+
\frac{\bm\tau}{2}.
}
\label{eq:grandspin}
\end{equation}
For the spherical hedgehog,
\begin{equation}
[H_{\rm h},\bm K^2]=0,
\qquad
[H_{\rm h},K_z]=0.
\end{equation}
Thus $K$ is a good quantum number, whereas $L$ by itself is not. It is useful first to combine pseudospin and flavor:
\begin{equation}
\frac12\otimes\frac12
=
0\oplus1.
\end{equation}
Define
\begin{equation}
\bm J
=
\frac{\bm\sigma}{2}
+
\frac{\bm\tau}{2},
\qquad
j=0,1.
\end{equation}
Then $\bm K=\bm L+\bm J$.
A grand-spin harmonic may be written as
\begin{equation}
\Omega^{KM}_{jL}(\hat{\bm r})
=
\sum_{m_L,m_j}
C^{KM}_{Lm_L,jm_j}\,
Y_{Lm_L}(\hat{\bm r})\,
\ket{j,m_j}_{\sigma\tau}.
\label{eq:grandharmonic}
\end{equation}

For $K=0$, the allowed channels are
\begin{equation}
(j,L)=(0,0),
\qquad
(j,L)=(1,1).
\label{eq:K0channels}
\end{equation}
For $K=1$, the allowed channels are
\begin{equation}
(j,L)
=
(0,1),\quad
(1,0),\quad
(1,1),\quad
(1,2).
\label{eq:K1channels}
\end{equation}

\subsection{Radial form of the zero-mode equation}

Define
\begin{equation}
A
=
(\hat{\bm r}\cdot\bm\sigma)\otimes\mathbb I_\tau,
\qquad
B
=
\mathbb I_\sigma\otimes(\hat{\bm r}\cdot\bm\tau).
\label{eq:ABdef}
\end{equation}
Using the standard identity
\begin{equation}
\bm\sigma\cdot\bm\nabla
=
A
\left(
\partial_r-\frac{\bm\sigma\cdot\bm L}{r}
\right),
\label{eq:siggrad}
\end{equation}
and $A^2=\mathbb I$, Eq.~\eqref{eq:reduced-zero} becomes
\begin{equation}
\left[
\partial_r
-
\frac{\bm\sigma\cdot\bm L}{r}
-
\eta\,m(r)\,AB
\right]\psi_\eta=0,
\qquad
m(r)\equiv\frac{\Phi(r)}{v_F}.
\label{eq:masterop}
\end{equation}
\subsection{Projection onto a fixed grand-spin sector}

Starting from the zero-mode equation
\begin{equation}
\left[
\partial_r
-\frac{\boldsymbol{\sigma}\cdot\mathbf L}{r}
-\eta\,m(r)\,AB
\right]
\psi_{\eta}(\mathbf r)
=0,
\label{eq:zero-master-angular}
\end{equation}
with
\begin{equation}
A=
(\hat{\mathbf r}\cdot\boldsymbol{\sigma})\otimes\mathbb I_\tau,
\qquad
B=
\mathbb I_\sigma\otimes
(\hat{\mathbf r}\cdot\boldsymbol{\tau}),
\qquad
m(r)=\frac{\Phi(r)}{v_F},
\end{equation}
we now separate the radial dependence from the angular, pseudospin, and
flavor degrees of freedom.
Since the spherically symmetric hedgehog Hamiltonian commutes with the
grand angular momentum operators,
\begin{equation}
[H_{\rm h},\mathbf K^2]=0,
\qquad
[H_{\rm h},K_z]=0,
\end{equation}
the zero-mode problem can be solved independently in each sector of
fixed $K$ and $M$.

For a given $(K,M)$, the wavefunction can be expanded in the complete
set of grand-spin harmonics compatible with these quantum numbers:
\begin{equation}
{
\psi_{\eta;KM}(\mathbf r)
=
\sum_{\alpha}
g^\eta_{\alpha K}(r)\,
\Omega_{\alpha}^{KM}(\hat{\mathbf r}).
}
\label{eq:grand-spin-expansion}
\end{equation}
Here $\alpha$ labels the different pairs $(j,L)$ that can combine to
give the same total grand spin $K$. Explicitly,
\begin{equation}
\Omega_{\alpha}^{KM}
\equiv
\Omega_{jL}^{KM}
=
\sum_{m_L,m_j}
C^{KM}_{Lm_L,jm_j}\,
Y_{Lm_L}(\hat{\mathbf r})\,
\ket{j,m_j}_{\sigma\tau},
\end{equation}
where
\begin{equation}
\mathbf J
=
\frac{\boldsymbol{\sigma}}{2}
+
\frac{\boldsymbol{\tau}}{2},
\qquad
j=0,1,
\end{equation}
and
\begin{equation}
\mathbf K=\mathbf L+\mathbf J.
\end{equation}

Because the grand-spin harmonics contain no radial dependence,
\begin{equation}
\partial_r\psi_{\eta;KM}
=
\sum_{\alpha}
g'_{\alpha K}(r)\,
\Omega_{\alpha}^{KM}.
\label{eq:radial-derivative-expansion}
\end{equation}

The remaining operators act within the finite-dimensional space of
grand-spin harmonics with fixed $(K,M)$. In particular,
\begin{equation}
(\boldsymbol{\sigma}\cdot\mathbf L)
\Omega_{\alpha}^{KM}
=
\sum_{\beta}
(S_K)_{\beta\alpha}
\Omega_{\beta}^{KM},
\label{eq:SK-action}
\end{equation}
where
\begin{equation}
{
(S_K)_{\beta\alpha}
=
\left\langle
\Omega_{\beta}^{KM}
\middle|
\boldsymbol{\sigma}\cdot\mathbf L
\middle|
\Omega_{\alpha}^{KM}
\right\rangle .
}
\label{eq:SK-matrix-elements}
\end{equation}

Similarly,
\begin{equation}
AB\,\Omega_{\alpha}^{KM}
=
\sum_{\beta}
(C_K)_{\beta\alpha}
\Omega_{\beta}^{KM},
\label{eq:CK-action}
\end{equation}
with
\begin{equation}
{
(C_K)_{\beta\alpha}
=
\left\langle
\Omega_{\beta}^{KM}
\middle|
AB
\middle|
\Omega_{\alpha}^{KM}
\right\rangle .
}
\label{eq:CK-matrix-elements}
\end{equation}

Substituting Eqs.~(\ref{eq:grand-spin-expansion}),
(\ref{eq:radial-derivative-expansion}),
(\ref{eq:SK-action}), and
(\ref{eq:CK-action}) into Eq.~(\ref{eq:zero-master-angular}) gives
\begin{align}
0
=
\sum_{\beta}
\Bigg[
g'_{\beta K}(r)
&-
\frac{1}{r}
\sum_{\alpha}
(S_K)_{\beta\alpha}
g_{\alpha K}(r)
-
\eta\,m(r)
\sum_{\alpha}
(C_K)_{\beta\alpha}\,
g_{\alpha K}(r)
\Bigg]
\Omega_{\beta}^{KM}.
\label{eq:projected-sum}
\end{align}

Since the $\Omega_{\beta}^{KM}$ form a linearly independent basis,
the coefficient of each basis state must vanish separately. Thus,
for every channel $\beta$,
\begin{equation}
g^\prime_{\beta K}(r)
-
\frac{1}{r}
\sum_{\alpha}
(S_K)_{\beta\alpha}
g_{\alpha K}(r)
-
\eta\,m(r)
\sum_{\alpha}
(C_K)_{\beta\alpha}
\,g_{\alpha K}(r)
=0.
\label{eq:component-radial-equations}
\end{equation}

Introducing the radial-vector notation
\begin{equation}
\mathbf g_K(r)
=
\begin{pmatrix}
g_{1K}(r)\\
g_{2K}(r)\\
\vdots
\end{pmatrix},
\end{equation}
the complete system takes the compact matrix form
\begin{equation}
{
\mathbf g_K'(r)
-
\frac{S_K}{r}\mathbf g_K(r)
-
\eta\,m(r)C_K\mathbf g_K(r)
=0.
}
\label{eq:masterradial}
\end{equation}

Thus,
\begin{equation}
{
S_K=
[\boldsymbol{\sigma}\cdot\mathbf L]_K,
\qquad
C_K=[AB]_K
}
\end{equation}
denote the matrix representations of
$\boldsymbol{\sigma}\cdot\mathbf L$ and $AB$, respectively, restricted
to the finite-dimensional subspace of grand-spin harmonics with fixed
$K$ and $M$.

The projection therefore converts the original partial differential
equation in $(r,\theta,\phi)$ into a finite system of coupled ordinary
differential equations in $r$:
\begin{equation}
{
\text{PDE in }(r,\theta,\phi)
\quad\longrightarrow\quad
\text{matrix radial ODE for }\mathbf g_K(r).
}
\end{equation}
Different values of $K$ do not mix because $K$ is conserved, whereas
different $(j,L)$ channels belonging to the same $K$ can in general mix
through the matrices $S_K$ and $C_K$.

\subsection{$K=0$ sector}

Choose the basis
\begin{equation}
\Omega_0
\equiv
\Omega^{00}_{0,0}
=
Y_{00}\ket{S},
\end{equation}
where the spin-flavor singlet is
\begin{equation}
\ket{S}
=
\frac{1}{\sqrt2}
\left(
\ket{\uparrow}_\sigma\ket{\downarrow}_\tau
-
\ket{\downarrow}_\sigma\ket{\uparrow}_\tau
\right),
\label{eq:singlet}
\end{equation}
and
\begin{equation}
\Omega_1
\equiv
\Omega^{00}_{1,1}.
\end{equation}
Explicitly,
\begin{equation}
\Omega_1
=
\frac1{\sqrt3}
\left[
Y_{11}\ket{T_{-1}}
-
Y_{10}\ket{T_0}
+
Y_{1,-1}\ket{T_{+1}}
\right].
\label{eq:K0triplet}
\end{equation}

In the ordered basis $(\Omega_0,\Omega_1)$,
\begin{equation}
A_0=
\begin{pmatrix}
0&-1\\
-1&0
\end{pmatrix},
\qquad
B_0=
\begin{pmatrix}
0&1\\
1&0
\end{pmatrix},
\end{equation}
and therefore
\begin{equation}
C_0=A_0B_0=-\mathbb I_2.
\label{eq:C0}
\end{equation}
Furthermore,
\begin{equation}
S_0=
\begin{pmatrix}
0&0\\
0&-2
\end{pmatrix}.
\label{eq:S0}
\end{equation}

Writing
\begin{equation}
\bm g_0(r)
=
\begin{pmatrix}
g(r)\\
h(r)
\end{pmatrix},
\end{equation}
Eq.~\eqref{eq:masterradial} becomes
\begin{align}
g'(r)+\eta m(r)g(r)&=0,
\label{eq:gK0}\\
h'(r)+\frac{2}{r}h(r)+\eta m(r)h(r)&=0.
\label{eq:hK0}
\end{align}

The solutions are
\begin{align}
g(r)
&=
g_0
\exp\left[
-\eta\int_0^r m(r')\,dr'
\right],
\label{eq:gsol}\\
h(r)
&=
\frac{h(r_0) r_0^2}{r^2}
\exp\left[
-\eta\int_{r_0}^r m(r')\,dr'
\right].
\label{eq:hsol}
\end{align}

For $\eta=+1$, $g(r)$ is finite at $r=0$ and decays as
\begin{equation}
g(r)\sim e^{-\Phi_0 r/v_F}
\end{equation}
for $r\to\infty$. It is therefore normalizable.

By contrast,
\begin{equation}
h(r)\rightarrow 0,\quad \text{as} \,\, r_0 \rightarrow 0.
\end{equation}
diverges. Hence the $(j,L)=(1,1)$ component does not provide a second
regular zero mode.

For $\eta=-1$, both independent solutions grow exponentially at large
$r$ and are non-normalizable. Therefore
\begin{equation}
{
n_+(K=0)=1,
\qquad
n_-(K=0)=0.
}
\label{eq:K0count}
\end{equation}

The unique regular zero mode in this sector is
\begin{equation}
{
\Psi_0(\bm r)
=
\mathcal N
\exp\left[
-\frac{\Phi_0}{v_F}
\int_0^r f(r')\,dr'
\right]
\ket{\rho_y=+}
\otimes
Y_{00}\ket{S}.
}
\label{eq:zeromode}
\end{equation}

\subsection{Callias index and higher winding}
\label{sec:index}

The explicit solution for the unit hedgehog shows that the monopole core supports a localized zero-energy fermionic state. To understand why this result is tied to the topology of the monopole, and to extend the argument to textures with arbitrary winding number, we use the Callias index theorem. The theorem relates the number of unpaired zero modes of the Dirac operator directly to the winding of the adjoint mass field at spatial infinity. It therefore provides a topological characterization of the zero modes that is independent of the detailed shape of the monopole core.\\

Let
\[
P_\pm=\frac{1\pm\Gamma}{2}
\]
denote the projectors onto the two \(\Gamma\)-sectors. Because of Eq.~\eqref{eq:grading},
\[
P_\pm H P_\pm=0.
\]
In a basis arranged according to the eigenvalue of \(\Gamma\), the Hamiltonian takes the off-diagonal form
\begin{equation}
H=
\begin{pmatrix}
0&D^\dagger\\
D&0
\end{pmatrix},
\qquad
\operatorname{ind}D=
\dim\ker D-\dim\ker D^\dagger .
\label{eq:Soffdiagonal}
\end{equation}
Our convention associates \(\ker D\) with the \(\rho_y=+1\) sector and is chosen such that the unit hedgehog in Eq.~\eqref{eq:zeromode} has positive index. Up to constant unitary factors,
\begin{equation}
D=
v_F\bm\sigma\cdot\bm\nabla
-
\bm\Phi(\mathbf r)\cdot\bm\tau .
\label{eq:SD}
\end{equation}

When the adjoint field approaches a finite magnitude at spatial infinity,
\[
|\bm\Phi(\mathbf r)|\rightarrow\Phi_0>0,
\]
the operator \(D\) satisfies the conditions of the Callias index theorem. The theorem expresses its index entirely in terms of the direction of the asymptotic mass field,
\begin{equation}
\operatorname{ind}D
=
\frac{1}{4\pi}
\int_{S^2_\infty}
d\vartheta\,d\varphi\,
\hat{\bm\Phi}\cdot
\left(
\partial_\vartheta\hat{\bm\Phi}
\times
\partial_\varphi\hat{\bm\Phi}
\right)
=
N .
\label{eq:SCallias}
\end{equation}
Thus the same winding number \(N\) that characterizes the monopole texture also determines the imbalance of zero modes between the two \(\Gamma\)-sectors.

For the elementary hedgehog, \(N=1\), and the theorem gives
\[
\operatorname{ind}D=1.
\]
This agrees with the explicit zero-mode solution obtained above and shows that the appearance of the unpaired mode is directly connected to the monopole winding. Smooth changes of the radial profile or of the detailed structure of the core leave the index unchanged as long as the asymptotic gap remains open, the winding number is preserved, and the spectral ( \(\Gamma\)) symmetry is maintained.

The index gives the difference between the numbers of zero modes in the two \(\Gamma\)-sectors. Additional zero modes may appear in pairs belonging to opposite sectors while leaving the index unchanged. For the elementary smooth profile considered here, the explicit solution shows that the minimal model contains a single unpaired zero mode.

The same argument applies directly to a texture with winding number \(N\). Equation~\eqref{eq:SCallias} gives
\[
\operatorname{ind}D=N,
\]
so the texture supports at least \(|N|\) unpaired zero modes, with the relevant \(\Gamma\)-sector determined by the sign of \(N\). In the minimal case, where these are the only zero modes, and \(k\) of them are occupied, the allowed charges are
\begin{equation}
Q=
e\left(
k-\frac{|N|}{2}
\right),
\qquad
k=0,1,\ldots,|N|.
\label{eq:Shighercharge}
\end{equation}
For odd \(|N|\), the allowed charges are half-odd-integer multiples of \(e\), while for even \(|N|\) they are integer multiples of \(e\). The winding number therefore determines the number and \(\Gamma\)-sector imbalance of the protected zero modes, whereas their occupation determines the charge carried by a particular many-body state.


\section{Fractional charge of the system in presence of point defects}
\label{sec:charge}

\subsection{Isolated defect}

We first consider an isolated point defect, as can occur in an infinite
system or in a system with an open boundary. The charge is defined relative
to the spectrally symmetric half-filled reference state through the
symmetrically ordered charge operator
\begin{equation}
\hat Q
=
\frac{e}{2}
\int d^3r\,
[\hat\Psi^\dagger(\mathbf r),\hat\Psi(\mathbf r)].
\label{eq:SQoperator}
\end{equation}
The symmetric ordering treats positive- and negative-energy states on equal
footing. Because every nonzero state at energy $E$ has a partner at $-E$,
their contributions cancel in the half-filled reference state.

The situation is different for an unpaired zero mode. If $a_0$ denotes the
fermionic operator associated with a single complex zero mode, its
contribution to the charge operator is
\begin{equation}
\hat Q_0
=
e\left(
a_0^\dagger a_0-\frac12
\right).
\label{eq:SQzero}
\end{equation}
The factor $1/2$ follows directly from the symmetric ordering,
\begin{equation}
\frac{e}{2}
\left(
a_0^\dagger a_0-a_0a_0^\dagger
\right)
=
e\left(
a_0^\dagger a_0-\frac12
\right).
\end{equation}

The zero mode can either be empty or occupied. These two possibilities give
\begin{equation}
n_0=0
\quad\Longrightarrow\quad
Q_0=-\frac{e}{2},
\qquad
n_0=1
\quad\Longrightarrow\quad
Q_0=+\frac{e}{2}.
\end{equation}
Thus an isolated hedgehog supporting a single unpaired zero mode can carry
a fractional charge $\pm e/2$ relative to the symmetric half-filled
reference~\cite{jackiwrebbi,goldstonewilczek}. The two charge states are
related by the exact $E\leftrightarrow -E$ spectral symmetry.

\subsection{Monopole - anti-monopole pair}

The situation is slightly different in a finite lattice with periodic
boundary conditions. The two $\Gamma$ sectors then contain the same total
number of states. In this case, the matrix $D$ in
Eq.~\eqref{eq:Soffdiagonal} is finite and square, and consequently
\begin{equation}
\dim\ker D
=
\dim\ker D^\dagger,
\qquad
\operatorname{ind}D=0.
\label{eq:Sfiniteindex}
\end{equation}
Hence a zero mode in one $\Gamma$ sector must be accompanied by a zero mode
in the opposite sector. In a periodic system this is naturally realized by
a hedgehog--antihedgehog pair. More generally, the compensating mode could
also occur at a physical boundary or in another spatial region.

Let $|M\rangle$ and $|A\rangle$ denote the localized modes associated with
a well-separated monopole and antimonopole. When their wavefunctions
overlap weakly, the two zero modes hybridize. In the basis
$\{|M\rangle,|A\rangle\}$, the low-energy Hamiltonian takes the form
\begin{equation}
H_{\rm pair}
=
t(d)
\begin{pmatrix}
0&1\\
1&0
\end{pmatrix},
\qquad
|\pm\rangle
=
\frac{|M\rangle\pm|A\rangle}{\sqrt2},
\qquad
E_\pm=\pm t(d),
\label{eq:Spairmodel}
\end{equation}
where the overlap decreases approximately as
\begin{equation}
t(d)\propto e^{-d/\ell}
\end{equation}
for a separation $d$ much larger than the localization length $\ell$.

The charge of the two low-energy modes is most clearly understood using the
same symmetric charge operator as above. If $a_M$ and $a_A$ denote the
operators associated with the two localized modes, the low-energy charge
operator is
\begin{equation}
\hat Q_{\rm pair}
=
e\left(
a_M^\dagger a_M-\frac12
\right)
+
e\left(
a_A^\dagger a_A-\frac12
\right).
\label{eq:Spaircharge}
\end{equation}
Thus each localized mode carries the same $-1/2$ subtraction that appears
for an isolated zero mode.

For a symmetric pair, the hybridized operators are
\begin{equation}
a_\pm
=
\frac{1}{\sqrt2}
\left(
a_M\pm a_A
\right).
\end{equation}
At neutrality, the negative-energy state is occupied while the
positive-energy state is empty,
\begin{equation}
\langle n_-\rangle=1,
\qquad
\langle n_+\rangle=0.
\end{equation}
This gives
\begin{equation}
\langle n_M\rangle
=
\langle n_A\rangle
=
\frac12.
\end{equation}
Substituting this into Eq.~\eqref{eq:Spaircharge}, the charge localized near
each core is
\begin{equation}
Q_M
=
e\left(
\frac12-\frac12
\right)
=
0,
\qquad
Q_A
=
e\left(
\frac12-\frac12
\right)
=
0.
\end{equation}
Therefore the neutral finite pair does not carry a separately localized
charge $\pm e/2$ at the two cores.

The fractional charge becomes visible when the occupation of the
near-zero-energy pair is changed. If both states are empty, then
$\langle n_M\rangle=\langle n_A\rangle=0$, and each core carries
$-e/2$. If both states are occupied, then
$\langle n_M\rangle=\langle n_A\rangle=1$, and each core carries
$+e/2$. Equivalently, adding one fermion to the neutral configuration by
occupying the positive-energy state adds a total charge $e$, which is
shared equally between the two well-separated and symmetry-related cores.

\begin{table}[h]
\caption{Occupation of the monopole--antimonopole near-zero-energy states.
The last column gives the corresponding core charges for a symmetric,
well-separated pair.}
\label{tab:Soccupations}
\begin{ruledtabular}
\begin{tabular}{ccc}
Occupation & Total charge relative to half filling & Local core charge\\
\hline
Both states empty & $-e$ & $-e/2$ at each core\\
Only $-t$ occupied & $0$ & $0$ at each core\\
Both states occupied & $+e$ & $+e/2$ at each core
\end{tabular}
\end{ruledtabular}
\end{table}

The same result can be seen from the spatial structure of the hybridized
states. For a symmetric pair, the positive- and negative-energy states have
the same probability density,
\begin{equation}
|\Psi_j^+(\mathbf r)|^2
=
|\Psi_j^-(\mathbf r)|^2
=
\frac12
\left(
|u_j(\mathbf r)|^2+
|v_j(\mathbf r)|^2
\right).
\label{eq:Ssvddensity}
\end{equation}
For the lowest-energy pair, $u_0$ and $v_0$ are localized near the two
opposite defect cores. Consequently, changing the occupation of one
hybridized state changes the total charge by $e$, while in the
large-separation and symmetric limit approximately $e/2$ of this added or
removed charge is localized near each core. The hedgehog--antihedgehog pair
therefore provides a finite periodic realization of the same zero-mode
fractionalization that, for an isolated defect, appears directly as the
$\pm e/2$ charge of a single unpaired zero mode.